\documentclass[showpacs, pra,twocolumn,preprintnumbers ,amsmath, amssymb, superscriptaddress, aps]{revtex4-2}
\usepackage{color}
\usepackage{amsmath,amssymb}
\usepackage{pifont}
\usepackage{amssymb}  
\usepackage{bbold}
\usepackage{float}
\usepackage{subfloat}

\usepackage[caption=false]{subfig}
\usepackage{tikz}
\usepackage{makecell}
\usepackage{subfig}
\usepackage{pifont}   
\usepackage{graphicx} 
\graphicspath{{Figures/}}
\usepackage{dcolumn}  
\usepackage{bm}       
\usepackage{multirow} 
\usepackage{placeins}
\usepackage[colorlinks]{hyperref}
\usepackage{mathtools}
\usepackage{appendix}

\def \be{\begin{align}}
\def \ee{\end{align}}
\def \bea{\begin{eqnarray}}
\def \eea{\end{eqnarray}}

\begin{document}
\title{Zero-energy states in graphene quantum dot with wedge disclination}
\author{Ahmed Bouhlal}
\affiliation{ Laboratory of Theoretical Physics, Faculty of Sciences, Choua\"ib Doukkali University, PO Box 20, 24000 El Jadida, Morocco}
\author{Ahmed Jellal}
\affiliation{ Laboratory of Theoretical Physics, Faculty of Sciences, Choua\"ib Doukkali University, PO Box 20, 24000 El Jadida, Morocco}
\affiliation{
Canadian Quantum  Research Center,
204-3002 32 Ave Vernon, BC V1T 2L7,  Canada}
\author{Nurisya Mohd Shah}
\affiliation{
Laboratory of Computational Sciences and Mathematical Physics, Institute for Mathematical Research (INSPEM), Universiti Putra Malaysia, 43400 UPM Serdang, Selangor, Malaysia}
\affiliation{
Department of Physics, Faculty of Science, Universiti Putra Malaysia, 43400 UPM Serdang, Selangor, Malaysia}	


\begin{abstract}
We investigate the effects of wedge disclination on charge carriers in circular graphene quantum dots subjected to a magnetic flux. Using the asymptotic solutions of the energy spectrum for large arguments, we approximate the scattering matrix elements, and then study the density of states. It is found  that the density of states shows several resonance peaks under various conditions. In particular, it is shown that the wedge disclination is able to change the amplitude, width, and positions of resonance peaks. 

\end{abstract}
\maketitle

\section{Introduction}
Graphene is the focus of much theoretical and experimental research due to its electronic and optical properties \cite{novoselov2004electric, chen2008intrinsic, geim2010rise, schedin2007detection}. Graphene has become a primary material for future nanoelectronic devices \cite{neto2009electronic}. Most of the research conducted so far is based on the mobility of electrons, which is determined by the diffusion of charged impurities \cite{martin2008observation, das2008monitoring}. The two graphene conduction and valence bands touch each other at two asymmetric points noted $(K, K')$ are called Dirac points. In the vicinity of these Dirac points, the energy dispersion is linear, which is the origin of graphene's unique electronic properties like high electrical conductivity \cite{novoselov2007electronic, neto2009electronic}. At normal incidence, the electrons transmit completely, which is due to a Klein tunneling effect, and it is therefore difficult to control them with external fields in the graphene \cite{young2009quantum, klein1929reflexion, katsnelson2006chiral}. However, it has been shown theoretically that it is still possible to create quasi-bound states (QBS) using external electric fields such as the magnetic field \cite{pereira2006confined, matulis2008quasibound, bardarson2009electrostatic}.

The notion of quantum dots is used in modern physics to study some physical phenomena. It is essential to the progress of quantum computing and spintronics. The reasons for choosing the application of graphene are twofold: on the one hand, because graphene is considered a two-dimensional zero energy gap system \cite{dora2009majorana} and on the other hand, because the quantum dots in graphene have the potential to be used as spin qubits  \cite{recher2010quantum}. The study of the possibility of electron confinement in graphene quantum dots is particularly important in this context \cite{ponomarenko2008chaotic, jacobsen2012transport}. The experimental activity will be based on the confinement of Dirac fermions in graphene quantum dots \cite{hamalainen2011quantum}. Due to the absence of a band gap, the confinement of charge carriers under the effect of an electrostatic potential using metallic gates is impossible in graphene.
The use of a magnetic flux in graphene quantum dots allows for better control and confinement of fermions in the presence of an electrostatic potential \cite{de2007magnetic}. It is well demonstrated that the magnetic flux is responsible for the kinematic angular momentum taking integer values, allowing the appearance of states that cannot be confined by the electrostatic potential alone \cite{pal2011electric}. The disk-shaped quantum dot has a circular symmetry from which the angle of incidence is a characteristic of the movement of electrons, knowing that the latter at their non-zero angular momentum are confined within the quantum dot \cite{bardarson2009electrostatic, calvo2011electrostatic}. Like graphene, it is well demonstrated that it is an ideal two-dimensional system where the energy dispersion is linear and has a density of states that cancels at zero energy \cite{neto2009electronic}.  In this work, we study electrostatic confinement in the presence of magnetic flux with wedge disclination.

Insulators and superconductors react significantly to topological defects and exhibit quasi-zero energy states around the defect \cite{de2014bound}.  The effect of a Coulombian electrostatic potential in the presence of a uniform magnetic field on the energy spectrum of graphene has been studied recently \cite{choudhari2014graphene}. Prior to the year 2014, two-terminal conductance was the only technique used to study the confinement of Dirac fermions in a graphene quantum dot \cite{bardarson2009electrostatic, heinl2013interplay, schneider2011resonant}. In 2014, Martin Schneider developed a technique based on density of states (DOS) analysis to study the possibility of confinement in a small graphene region \cite{schneider2014density}. These studies are extended by examining how the information of the confinement signature has been deduced by the analysis of the density of states, and in parallel, the effect of the gap energy on the information of the system has also been studied \cite{bouhlal2021density}. We wish to consider in this context the simple disinclination via the construction of Volterra \cite{furtado1994landau}.Our goal in this paper is to investigate the properties of a graphene monolayer in which the crystal symmetry is locally modified by the conical geometry mentioned above.

We consider a circular graphene quantum dot of radius $R$ surrounded by a circular ring of undoped graphene of external radius $L$ bound by a metallic contact in the presence of a magnetic field with a wedge disclination $n = 0, \pm1, \pm2$. By restricting ourselves to the asymptotic behavior of Hankel functions for large arguments, we determine the scattering matrix. DOS is calculated as a function of magnetic flux $\Phi$, applied electrostatic potential $V_0$, and wedge disclination $n$. 
 We numerically analyze the DOS under various conditions of the physical parameters. This allows us to show different oscillatory behaviors and to find resonances. We show that the DOS peaks strongly depend on the momentum quantum number $ m $. During the scanning of the gate voltage, the zero energy bound states cause resonant peaks that become narrow if the size $L$ of the undoped graphene sheet containing the quantum dot increases, see Fig. \ref{fig1}. 
Subsequently, we discover that  $n$ acts by changing the resonance peaks. Indeed, this behavior contrasts strongly with that of $n = 2$ (square defect), which moves to the right from its initial position where $n = 0$ (defect-free), and then the resonance becomes more and more acute. In the opposite case, where $n = -2$ (octagon defect), we observe the creation of a new peak that corresponds to $m = 5/2$.

The paper is organized as follows. In section \ref{theo}, we establish a theoretical model describing electrostatically confined charge carriers in geometry obtained by Volterra construction. We calculate the scattering matrix by restricting to the asymptotic behavior of Hankel functions in section \ref{SMA}, which are obtained in the solution of the energy spectrum. Using the scattering matrix, we explicitly determine the associated DOS in section \ref{DOS}. In section \ref{res}, we numerically analyze the DOS and show it exhibits a set of resonance peaks. Under various conditions, the effects of magnetic flux and the wedge $ n $ on these peaks will be investigated. Finally, we conclude our results.

\section{Theoretical model} \label{theo}
The configuration we study consists of a graphene quantum dot (QD) connected to a ring-shaped metal contact, and the two are separated by an undoped intrinsic graphene layer.  
\begin{figure}[h]
	\centering 
	\includegraphics[scale=0.5]{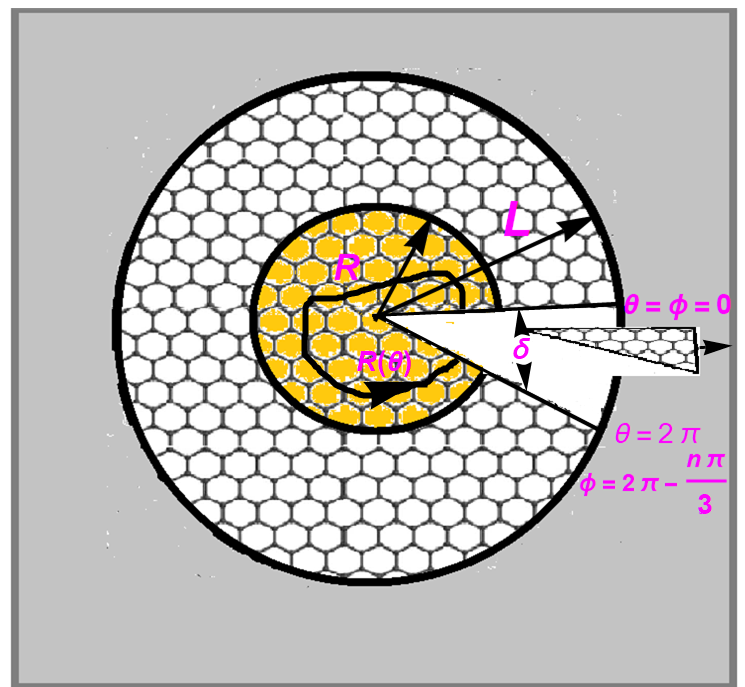}	
	\caption{(color online) 
		Graphene quantum dots of radius $R$ surrounded by an undoped intrinsic graphene sheet in the form of a ring of outer radius $L$ and coupled to source and drain reservoirs with a corner of angle $\theta$ removed.  $R(\theta)$ is the curve along which the spinor is rotated.}\label{fig1}
\end{figure} 
The Hamiltonian of a graphene monolayer that describes the motion of a single electron in the presence of an axially symmetric external potential $V(r) $ at low energy is 
\begin{equation} \label{hamiltonian}
H=v_{F}\left({\tau_z} \sigma_{x} p_{x}+\sigma_{y} p_{y}\right)+V(r) \mathbb{I}
\end{equation}
where $p=(p_x,p_y)$ is the  momentum operator, $v_F = 10^6$ m/s is the Fermi velocity, $\sigma_i, \tau_i$ {($ i=x,y,z $)} are Pauli matrices denoting the sublattice and valley degrees of freedom in the basis of the two sublattices of $A$ and $B$ atoms, respectively. The potential barrier $V(r)$ is  defined as follows
\begin{equation}\label{ev2}
V(r)=
\left\{%
\begin{array}{lll}
0, & {R}<r<{L}\\
- \hbar v_F V_0,&  r<{R}  \\
- \hbar v_F V_\infty,&  r>{L}
\end{array}%
\right.
\end{equation}
with the parameters $V_0$ and $V_\infty$ are chosen to be 
positive because the two regions outside the ring of radius $R$ and $L$ are doped with electrons. The highly doped lead graphene is modeled by taking the limit $V_{\infty} \to \infty$. The Hamiltonian \eqref{hamiltonian} acts on the spinor with four components
\begin{equation} \label{psi1}
\Psi(\vec{r})=\left(\psi_{A+}(\vec{r}) \psi_{B+}(\vec{r}) \psi_{A-}(\vec{r}) \psi_{B-}(\vec{r})\right)^{T}
\end{equation}
where the valley indices $+(-)$ refer to the two Dirac points $K(K')$ in the Brillouin zone. For a two dimensional graphene, the angular boundary condition for a Dirac spinor when it follows a closed path is given by
\begin{equation} \label{psi2}
\Psi(r, \theta=2 \pi)=e^{i \pi {\tau_z} \sigma_{z}} \Psi(r, \theta=0)
\end{equation}
and we will see that the presence of the wedge disclination modifies the angular boundary conditions in  \eqref{psi2}. Indeed, it is known that the deformation in a honeycomb lattice falls in the description at the continuum boundary undergoes a pseudo-gauge field coupled to the electron pulse by a vector potential \cite{vozmediano2010gauge, guinea2010energy, ghaemi2012fractional}. 
As shown in Fig. \ref{fig1}, the Volterra construction can be used to introduce the wedge disclination by removing a sector of $\delta=n \pi/3$. The integer $n$ represents the type of disclination in the honeycomb lattice, with $n=-2$ representing an isolated octagon defect, $n=-1$ representing an isolated heptagon defect, $n=0$ representing an isolated defect in the no defect case, $n=1$ representing an isolated pentagon defect, and $n=2$ representing an isolated square defect. 
  As a result, the wave function  \eqref{psi2} becomes 
 \begin{equation} \label{psi3}
 \Psi(r, \theta=2 \pi)=-e^{i 2 \pi[1-(n / 6)] {\tau_z}  \sigma_{z} / 2} \Psi(r, \theta=0).
 \end{equation}
Exchanging the $K$ and $K'$ blocks in the Hamiltonian using the unitary transformation $e^{i \pi \sigma_{y} \tau_{y} /2}$ yields the angular boundary condition for any number $ n $
 \cite{lammert2000topological, ruegg2013bound}
 \begin{equation} \label{psi4}
 \Psi(r, \theta=2 \pi)=-e^{i 2 \pi\left[-\frac{n}{4} \sigma_{y} \tau_{y}+(1-(n / 6)) \frac{1}{2} {\tau_z} \sigma_{z}\right]} \Psi(r, \theta=0)
 \end{equation}
 For a local gauge transformation,  \eqref{psi4} can be written in two singular gauge transformations
 \begin{equation}
 \Psi(r, \theta=2 \pi)=F(\phi) G_{n}(\theta) \Psi(r, \theta=0)
 \end{equation}
 where  $F(\phi)=e^{i \frac{\phi}{2} {\tau_z} \sigma_{z} }$  and  $G_{n}(\theta)=e^{i \frac{n \theta}{4} \tau_{y} \sigma_{y}}$, as shown in  Fig. \ref{fig1} with $\theta= \frac{\phi}{1-\frac{n}{6}}\in  [0,2 \pi]$.

Let us now consider an external magnetic flux $\Phi$ applied to the quantum dot (QD) 
\begin{equation}
\vec A = 
\frac{\hbar \Phi}{\Omega_n e \Phi_0 r} \vec{e}_\theta
\end{equation}
where $\Phi_0=h/e$ is the flux quantum and $\vec{e}_\theta$ is the unit vector along the azimuthal direction and the wedge disclination is described by $\Omega_n=1-\frac{n}{6}$ \cite{nazarov2009quantum} with $n=0,\pm1,\pm2$. For this, 
we replace the momentum $\vec{p}$ by the conical momentum $\vec{p}+e \vec{A}$ in the Hamiltonian \eqref{hamiltonian} and  use the transformations of $F(\phi)$ and $G_{n}(\theta)$. This process yields  
\begin{equation}
\tilde{H}(r, \theta)=F^{\dagger} G_{n}^{\dagger} H G_{n} F
\end{equation}
such that the Hamiltonian \eqref{hamiltonian} is now given by
\begin{align}
\tilde{H}(r, \theta)=& \hbar v_{F}\left(k_{r}-\frac{i}{2 r}\right)  \sigma_{x} \\
&+\hbar v_{F}\left(k_{\theta}+\frac{\Phi_i}{\Omega_{n} r}+\frac{n}{4 \Omega_{n} r} \tau_{z}\right) \sigma_{y}+V(r) \mathbb{I}\notag
\end{align}
where $k_{r}=-i \frac{\partial}{\partial r}$, $k_{\theta}=-\frac{i}{r \Omega_{n}} \frac{\partial}{\partial \theta}$ and $\Phi_i=\frac{\Phi}{\Phi_0}$ is the 
dimensionless flux. 
It is easy to verify that the total angular momentum $J_z=L_z+\frac{\hbar}{2}\sigma_z$ commutes with the Hamiltonian \eqref{hamiltonian}. As a result, the eigenfunctions can be 
separated as
\begin{equation}\label{e2}
\Psi(r,\theta)=e^{i m \theta}
\left(%
\begin{array}{c}
\chi_A(r) \\
\chi_B(r) \\
\end{array}%
\right)
\end{equation}
with $m=\pm1/2,\pm3/2 \cdots $ are eigenvalues  of $J_z$.

We solve the eigenvalue equation $H \Psi_{\tau} = E \Psi_{\tau}$ in the three regions: $0 < r < R$, $R < r < L$ and $r > L$.
After some algebras, we obtain two coupled radial equations
\begin{align}
&	\label{equ1}
\left[\frac{\partial}{\partial r}+\frac{1}{2 r}+\frac{1}{r \Omega_{n}}\left(m+\Phi_i{+ \frac{n \tau}{4}}\right)\right] \chi_{B}={(\epsilon-V_i)\chi_A}
\\
&\label{equ2}
\left[-\frac{\partial}{\partial r}-\frac{1}{2 r}+\frac{1}{r \Omega_{n}}\left(m+\Phi_i{+ \frac{n \tau}{4}}\right)\right] \chi_{A}={(\epsilon-V_i)\chi_B}
\end{align}
where 
dimensionless parameters $\epsilon=\frac{E}{\hbar v_F}$ and $V_i=\frac{V(r)}{\hbar v_F}$ have been defined. 
We look for a second order equation for one component and reusing one of (\ref{equ1}-\ref{equ2}) to find the other component of the wave function. Now, injecting \eqref{equ1} into  \eqref{equ2} 
to end up with 
\begin{equation}\label{e9}
\left[\rho^2 \frac{\partial^2}{\partial \rho^2}+\rho
\frac{\partial}{\partial \rho}+  \rho^2 - \left(\mu -\frac{1}{2}\right)^2
\right]\chi_A(\rho)=0
\end{equation}
where we have set  $\mu=\frac{1}{ \Omega_{n}}\left(m+\Phi_i{+ \frac{n \tau}{4}}\right)$ and $\rho=k r$ with $k=\epsilon-V_i$ is the wave number.
 \eqref{e9} can be reformulated as the Bessel differential equation. In general, the solution is a linear combination of the Hankel function of first $H^{+}_n\left(\rho\right)$ and of second $H^{-}_n\left(\rho\right)$ kinds. Then we combine them  to get the proper eigenspinors
\begin{equation} \label{eq:psiref}
\psi_{k,\mu}^{\pm}(r) =e^{i\left(\mu-\frac{1}{2}\right)\theta} \sqrt{\frac{k}{4\pi}}\begin{pmatrix}
H^\pm_{\mu-1/2}(k r) \\
i~\mathrm{sign}(\mu)e^{i\theta}H^{\pm}_{\mu+1/2}(k r)
\end{pmatrix}
\end{equation}
 which describe incoming (-) (propagating from $r = 0$) or outgoing (+)  (propagating from $r =\infty $) for circular waves. Note that, $H_{n}^{(\pm)}$ are linear combinations of Bessel $J_n$  and Neumann $Y_n$, namely $H_{n}^{(\pm)}=J_{n} \pm i Y_{n}$.
We specify the value of  $k$ which differs from one region to another where the solutions \eqref{eq:psiref} apply. For $r<R$: $k \equiv k_{0}=\epsilon +V_{0}$, for $R<r<L$: $k=\epsilon$ and for $r>L$: $k \equiv k_{\infty}=\epsilon+V_{\infty}$.
Recall that the half-integer Bessel function are valid, i.e., $Y_{1/2}(x)= - J_{-1/2}(x)= -\sqrt{\frac{2}{\pi x}} \cos x$, $Y_{-1/2}(x)=J_{1/2}(x)=\sqrt{\frac{2}{\pi x}}\sin x$. It is clearly seen that the function $\frac{\cos(kr)}{\sqrt{r}}$ is divergent at the origin.
Therefore for $\mu=0$, we have
\begin{equation} \label{eq:psirefflux0}
\psi_{\kappa,0}^{\pm}(r) =\frac{e^{\pm i \kappa r}}{\sqrt{8 \pi^2 r}}\begin{pmatrix}
\pm e^{- i \theta } \\
1
\end{pmatrix}.
\end{equation}
At the center of the quantum dot ($R \to 0$) and for the wave function to be regular at $r=0$, we have the solution inside the quantum dot ($r<R$)
\begin{equation} \label{eq:psidot}
\psi_{k,\mu}(r) =e^{i\left(\mu-\frac{1}{2}\right)\theta} \sqrt{\frac{k_0}{4\pi}}\begin{pmatrix}
J_{\mu-1/2}(k r) \\
i~\mathrm{sign}(\mu)e^{i\theta}J_{\mu+1/2}(k r).
\end{pmatrix}
\end{equation}

For the region $R<r<L$, the
eigenspinors can be written as a linear combination of the
two solutions of  \eqref{eq:psiref}
\begin{eqnarray}
\psi_2(r)&=&a_{\mu} \psi_{k,\mu}^-(r)+b_{\mu} \psi_{k,\mu}^+(r)
\end{eqnarray}
with $a_\mu$, $b_\mu$ being arbitrary constants.
Inside the circular ring  ${R}<r<{L}$ and at zero energy ($\epsilon=0$), the radial components have the forms
\begin{equation}
\label{psinoenergy}
\chi_A(r)= a_{+} r^{\mu-\frac{1}{2}}, \qquad \chi_B(r)= a_{-} r^{-\mu-\frac{1}{2}}.
\end{equation}
To avoid divergence, we require the constraints $a_{+}=0$ for $\mu>0$ and $a_{-}=0$ for $\mu<0$.

We use the asymptotic behavior of Hankel functions in the limit of a heavily doped lead where $k_{{\infty}} L \gg 1$. 
That is
\begin{equation}
H^{(\pm)}_n(x)\approx (2/\pi x)^{1/2} e^{\pm i(x-n\frac{\pi}{2}-\frac{\pi}{4})}.
\end{equation}
As a results the eigenspinors $\psi_1(r)$  inside the quantum dot ($r<R$) and $\psi_3(r)$ outside the quantum ring ($r>L$)  can be simplified to
\begin{eqnarray}
\psi_1(r)&=& \sqrt{k_0} \begin{pmatrix}
J_{\mu-1/2}(k_0 r) \\
J_{\mu+1/2}(k_0 r)
\end{pmatrix}\\
\psi_3(r)&=&c_\mu \frac{e^{ - i k_{\infty} r}}{\sqrt{r}} \begin{pmatrix}
1 \\
- 1
\end{pmatrix} + d_\mu \frac{e^{ i k_{\infty} r}}{\sqrt{r}} \begin{pmatrix}
1 \\
1
\end{pmatrix}
\end{eqnarray}

\section{Scattering matrix}\label{SMA}
The coefficients $a_{\mu}(\epsilon)$, $b_{\mu}(\epsilon)$, $c_{\mu}(\epsilon)$ and $d_{\mu}(\epsilon)$ can be determined using  the continuity of the wave function at $r = R$ and $r = L$ and the regularity at $r=0$. The first two coefficients are linked
via the scattering matrix $ S_{\mu}(\epsilon) $ 
\begin{align}
&\label{eq:scatt}
{d}_{\mu}(\epsilon) = \mathcal{S}_{\mu}(\epsilon) {c}_{\mu}(\epsilon)\\
&\label{S}
S_{\mu}(\epsilon) = - \frac{\det D^{(-)}}{\det D^{(+)}}
\end{align}
where the  matrices $D^{(\pm)}$ are given by
\begin{widetext}
\begin{equation}
\label{D12}
D^{(\pm)}=
\begin{pmatrix}
0 & \sqrt{k} H^{(+)}_{\mu-\frac{1}{2}} (k R) &\sqrt{k} H^{(-)}_{\mu-\frac{1}{2}} (k R) & \sqrt{k_0} J_{\mu-\frac{1}{2}}(k_0 R) \\
0 & \sqrt{k} H^{(+)}_{\mu+\frac{1}{2}} (k R) & \sqrt{k} H^{(-)}_{\mu+\frac{1}{2}} (k R) & \sqrt{k_0} J_{\mu+\frac{1}{2}}(k_0 R) \\
\frac{e^{ \mp i k_{\infty} L}}{\sqrt{L}} & - \sqrt{k} H^{(+)}_{\mu-\frac{1}{2}} (k L) & - \sqrt{k} H^{(-)}_{\mu-\frac{1}{2}} (k L) & 0 \\
\mp \frac{e^{ \mp i k_{\infty} L}}{\sqrt{L}} & - i \sqrt{k} H^{(+)}_{\mu+\frac{1}{2}} (k L) & - i \sqrt{k} H^{(-)}_{\mu+\frac{1}{2}} (k L) & 0
\end{pmatrix}.
\end{equation}
\end{widetext}
For small distances, we have
\begin{align}
&J_{n}(x)\sim \frac{1}{n!} \left(\frac{x}{2}\right)^{n}\\
&
Y_{n}(x) \sim \left\{\begin{array}{c} -\frac{\Gamma(n)}{\pi}\left(\frac{2}{x}\right)^n,~~n>0\\ \frac{2}{\pi}\ln\left(\gamma_E\frac{x}{2}\right),~~n=0
\end{array}  \right.
\end{align}
with $\ln\left(\gamma_E\right)=0.577\cdots$ is defined as an Euler constant. Indeed, for  small $\epsilon$, we can develop the scattering matrix as a function of $k$ in the internal region  ($R <r <L$). 
 Then, we write
\begin{align}
\mathcal{S}_{\mu}(\epsilon) = e^{-2 i k_{\infty} L + i |\mu| \pi}
\left[ \mathcal{S}_{\mu}^{(0)} + k L \mathcal{S}_{\mu}^{(1 )}
+ {\cal O}(\epsilon^2) \right]
\label{scat}
\end{align}
such that
\begin{equation}
\mathcal{S}_{\mu}^{(0)}=\frac{1+i \mathcal{J}_{\mu} R_L^{2|\mu|}}{1-i \mathcal{J}_{\mu} R_L^{2|\mu|}}
\end{equation}
and  we used the abbreviation 
\begin{equation}
{\cal J}_\mu=\frac{J_{|\mu|+\frac{1}{2}}\left( k_0 {R}\right)}{J_{|\mu|-\frac{1}{2}}\left( k_0 R\right)}
\end{equation}
where $R_L=\frac{R}{L}$.
We distinguish two cases for $S_{\mu}^{1}$. Indeed, in the absence of magnetic flux and in the defect-free case ($\phi=0, n=0$), we have 
	\begin{align}
	\label{eq:calS1j}
	\mathcal{S}^{(1)}_{\mu}=&
	\displaystyle
	-\frac{2 i }{2|\mu|-1} {\cal S}^{(0)}_\mu\\
	&
	+
	\frac{8 i |\mu| +2 i[(2|\mu|+1){\cal J}_\mu^2-(2|j|-1)]R_L^{2|\mu|+1} }{(4 |\mu|^2 - 1)(1 - i {\cal J}_j R_L^{2|\mu|})^2}\notag
	\end{align}
	as well as for $|\mu|=\frac{1}{2}$
	\begin{equation}
	\label{eq:S1j12}
	\mathcal{S}^{(1)}_{\pm 1/2}=
	\frac{i  (1-R_L^2)-2 i{\cal J}_{\frac{1}{2}}^2 R_L^2  \ln (R_L)}{(1-i{\cal J}_{\frac{1}{2}}R_L)^2}.
	\end{equation}
In the presence of magnetic flux and wedge disclination, we find that the results of \eqref{eq:calS1j} remain valid for $\mu\neq0$.
%
We compute the scattering matrix $S_0(\epsilon)$ for the case $\mu=0$. We show that it is constant and independent of $V_0$, such as
     	\begin{equation}\label{s0}
     S_0=e^{-2i(k_{\infty}-k_0)R} e^{-2i(k-k_{\infty})(R-L)}.
     \end{equation}
%

\section{Density of states}\label{DOS}

Using the scattering matrix for low energy \eqref{scat}, we will calculate the density of states at zero energy in terms of the potential $V(r)$. The local density of states (LDOS) $\rho(r,\epsilon)$ is given by the derivative of the scattering matrix  $S(\epsilon,V(r))$ \cite{langer1961friedel,buttiker1993capacitance,buttiker1994current}
\begin{equation}
\rho(r,\epsilon) = - \frac{1}{2 \pi i  } \mbox{Tr}\, { S_\mu(\epsilon,V(r))}^{\dagger} \left( \frac{\delta { S_\mu(\epsilon,V(r))}}{\delta V(r)}\right).
\end{equation}
We calculate the total density of states (DOS) of the circular quantum dot by integrating $\rho(r,\epsilon)$ in region $r<L$ \cite{schneider2014density}. This yields to
\begin{equation}
\rho_{dot}(\epsilon) = -\frac{1}{2 \pi i }
\int_{r < L} \mbox{Tr}{ S_\mu(\epsilon,V(r))}^\dagger \left( \frac{\delta { S_\mu(\epsilon,V(r))}}{\delta V(r)}\right)~dr.
\label{eq:nuWS}
\end{equation}
We can now calculate $\rho_{\rm dot}$ at zero energy using \eqref{eq:nuWS}, which gives
\begin{equation}
\label{eq:deltanu}
\rho_{dot} = \frac{1}{2 \pi i \hbar v_F} \sum_{\mu}
~ \mathcal{S}^{(0)*}_\mu \left[ \frac{\partial \mathcal{S}^{(0)}_\mu}{\partial V_0}
+L \mathcal{S}^{(1)}_\mu \right].
\end{equation}
It is worth noting that DOS is related to the point conductance \cite{buttiker2000time, schneider2014density} and is dependent on the Wigner-Smith delay  \cite{smith1960lifetime}.

Taking into account the continuity of the eigenspinors \eqref{eq:psidot} and \eqref{psinoenergy} at $r = R$ for zero energy, the resonance condition is well established
\begin{equation}
\label{eq:resonanceposition}
J_{|\mu|-1/2}(V'_0 {R})=0
\end{equation}
where $V'_0=V_0$. At resonance, one can write
\begin{equation}
{\cal J}_{\mu}\approx \frac{1}{R(V'_0 - V_0)}.
\end{equation}
For zero energy, we show that the DOS has a Lorentzian variation on $V_0$. 
 When there is no magnetic flux ($\phi =0$) and $\mu| \neq 1/2$,
DOS takes the form at zero energy 
\begin{equation}
\label{eq:deltanures}
\rho_{dot} = \frac{4 R |\mu|}{\pi \hbar v_F (2|\mu|-1)}
\frac{\Gamma}{4 R^2 (V_0 - V'_0)^2 + \Gamma^2} 
\end{equation}
while for $|\mu| = 1/2$, we have
\begin{equation}
\rho_{dot} = \frac{2 R}{\pi \hbar v_F}
\left(1 - \ln R_L \right)
\frac{\Gamma}{4 R^2 (V_0 - V'_0)^2 + \Gamma^2}
\end{equation}
where the resonance width is given by
\begin{equation}
\label{eq:Gamma}
\Gamma=2\left( R_L\right)^{ 2|\mu|}
\end{equation}
It is worth noting that \eqref{s0} demonstrates that there is a constant that is independent of the gate voltage $V_0$ as well as a non-resonant contribution to DOS $\rho_{dot}$. 
In order to study the possibility of electrostatic confinement in the graphene quantum dot reflected in DOS, we numerically analyze the analytical results obtained above. The next paragraph presents the studies on the effect of wedge disclination on the zero energy states in quantum-circular graphene.

\section{Numerical results and discussions} \label{res}
\begin{figure}[h]\centering
	\includegraphics[width=0.5\linewidth]{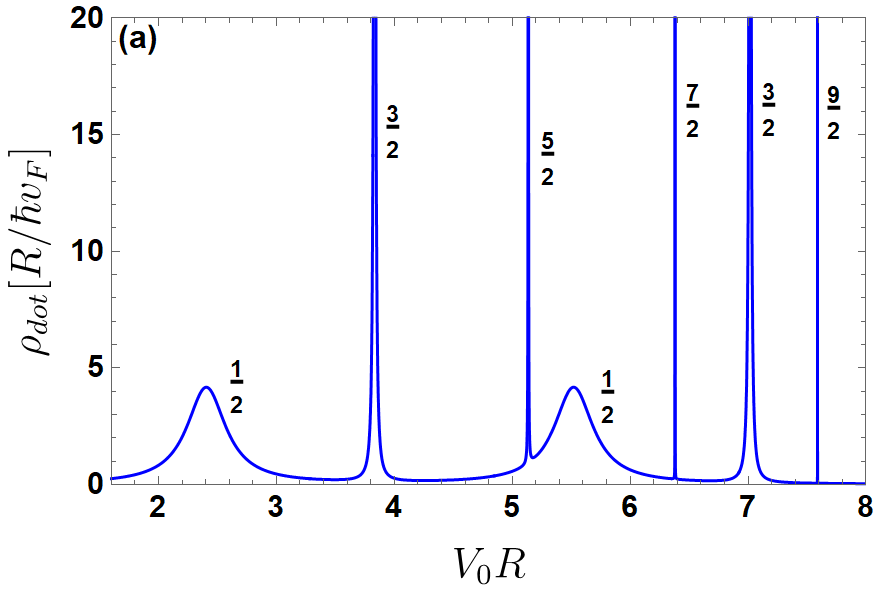}\includegraphics[width=0.5\linewidth]{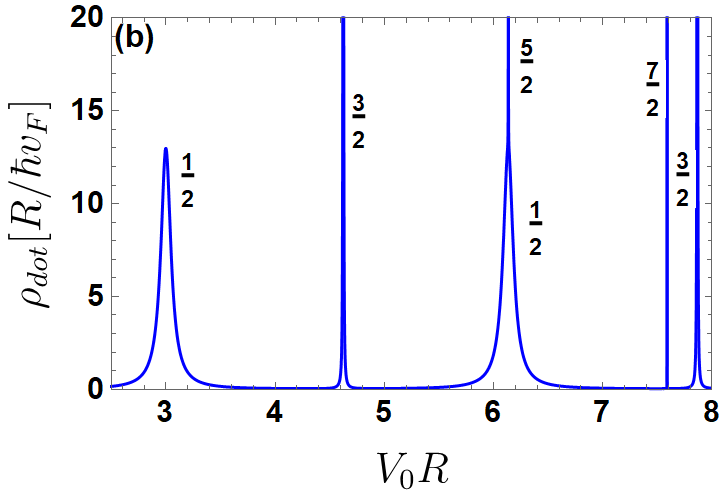}
	\includegraphics[width=0.5\linewidth]{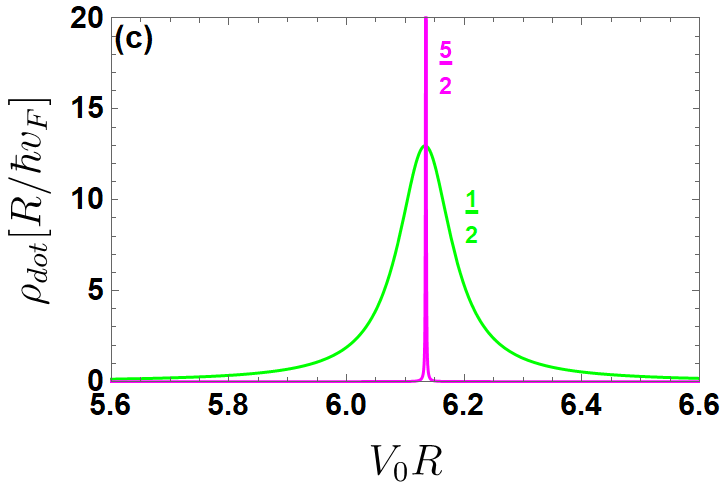}\includegraphics[width=0.5\linewidth]{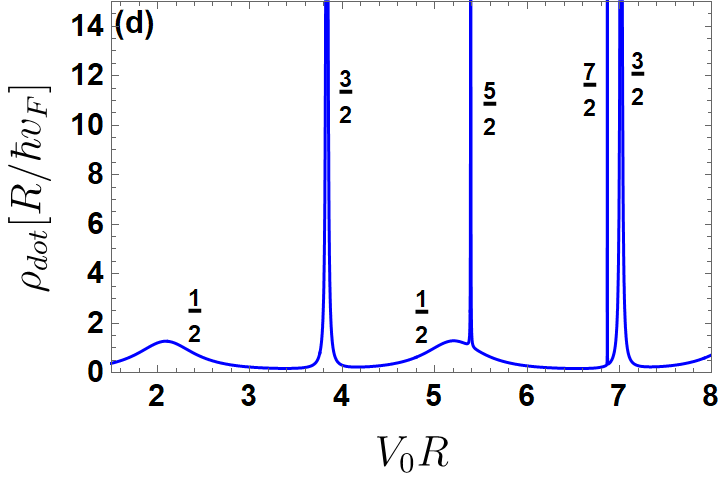} 
	\includegraphics[width=0.5\linewidth]{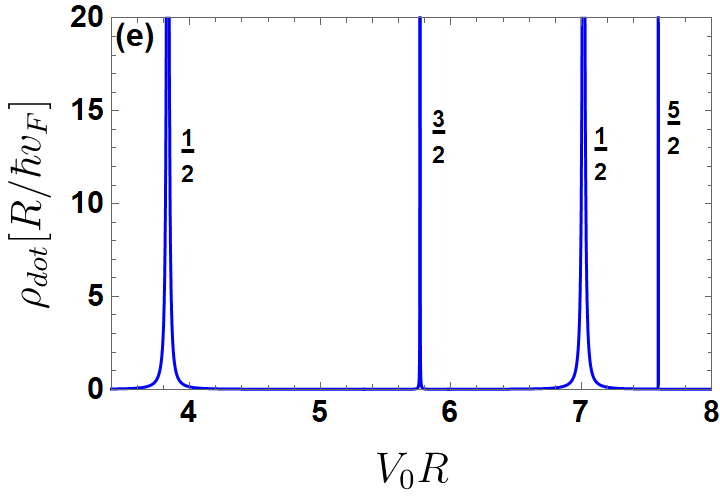}\includegraphics[width=0.5\linewidth]{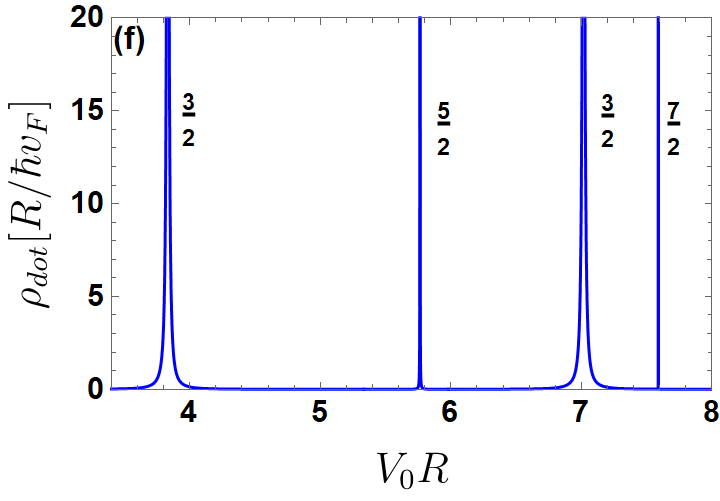}\\
	\includegraphics[width=0.5\linewidth]{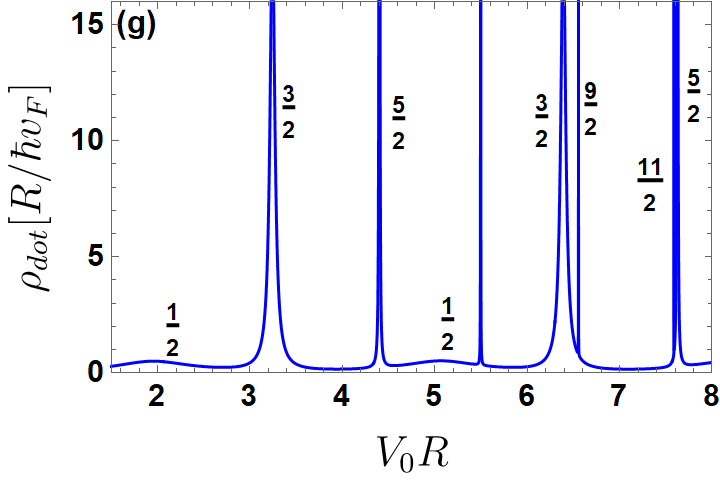}\includegraphics[width=0.5\linewidth]{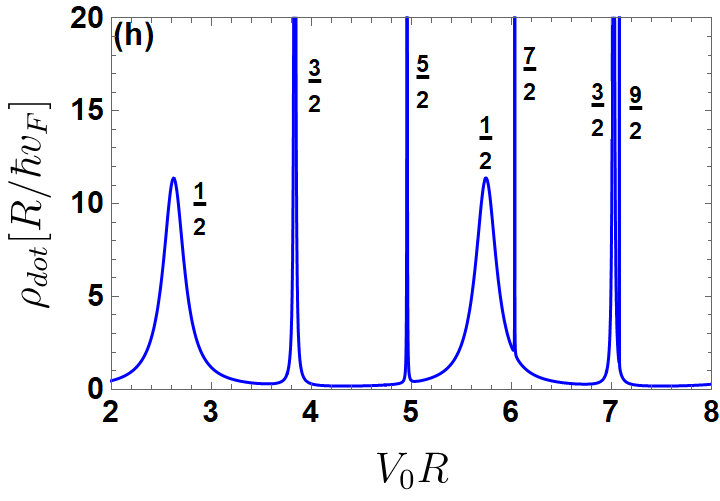}\\
	\includegraphics[width=0.5\linewidth]{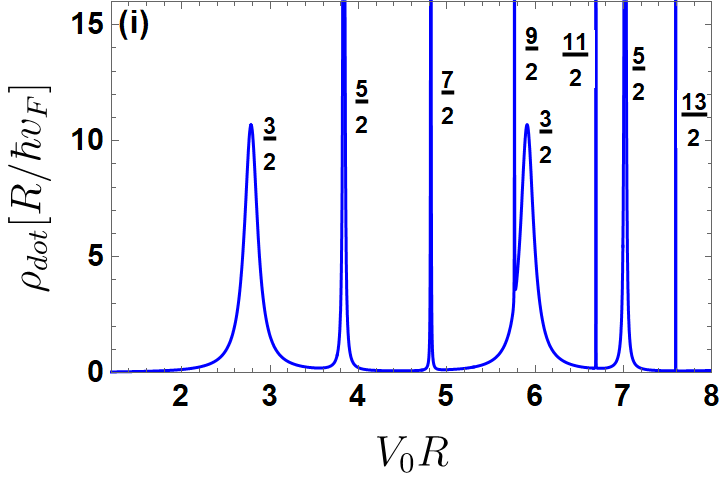}\includegraphics[width=0.5\linewidth]{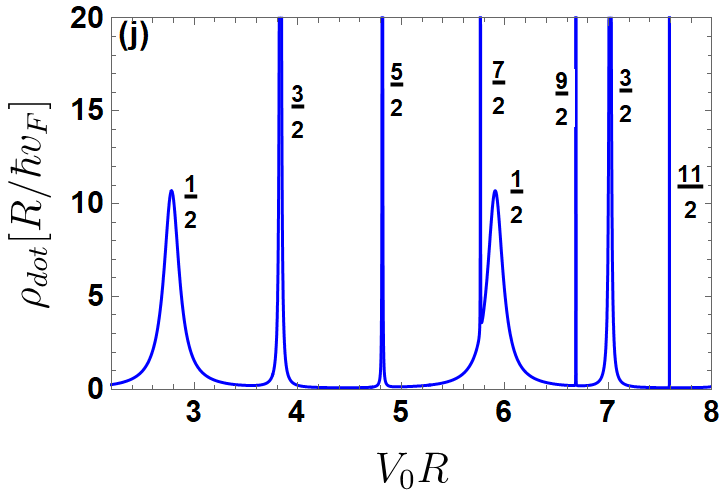}
	\caption{(color online) 
		DOS as a function of gate voltage $R V_0$ at $\epsilon =0$ with $R/L = 0.2$,   $\Phi_i=0$ for  $(n,\tau)$ such that (a): 0, (b): $(1, 1)$, (d): $(1, -1)$, (e): $(2, 1)$, (f): $(2, -1)$. (g): $(-1, 1)$. (h): $(-1, -1)$, (i): $(-2, 1)$, (j): $(-2, -1)$. Here  (c) is a zoom of (b) for  $5.6<V_0 R<6.6$ and resonances are labeled by angular momentum $|m|= 1/2, \cdots,  13/2$.    }
	\label{fig01}
\end{figure}
We start at this point with the numerical analysis of the results obtained analytically. First, we investigate the effect of ratio $R_L=\frac{R}{L}$, magnetic flux $\Phi_i$, and wedge disclination index $n$ values on the density of states of a quantum dot in graphene surrounded by an undoped hands-on graphene region with zero energy $\epsilon=0$ and coupled to source and drain reservoirs. We analyze the possibility of confining electrons in the quantum-circular of monolayer graphene by evaluating the DOS of such a circular quantum dot defined by a gate in the presence of two conductors. Indeed, because the parameters of our theory are $\mu=\frac{1}{ \Omega_{n}}\left(m+\Phi_i{+ \frac{n \tau}{4}}\right)$, then we choose to numerically analyze the DOS versus the gate voltage $R V_0$ under suitable conditions of the physical parameters characterizing our system. 
%


DOS as a function of $ V_0R$ is shown in Fig. \ref{fig01} by varying the values of the wedge disclination index $n = 0, \pm1, \pm2$, and valley $\tau=\pm1$. Recall that each value of $n$ represents a different type of default, such as $n= 0$ for a defect isolated in the absence of a defect (a), $n= 1$ for a defect isolated in a pentagon (b), $n= 2$ for a defect isolated in a square (c), $n= -1$ for a defect isolated in a heptagon (d), and $n= -2$ for a defect isolated in an octagon (e).   
We notice that the DOS presents resonance peaks with oscillatory behavior that are marked by their angular momentum $m$. For $n=0$, we notice that the position of the resonances, as well as the scale of the amplitude and width, are in good agreement with the results obtained in the literature
 \cite{bardarson2009electrostatic, titov2010charge, schneider2014density, bouhlal2021density}. 
 We find that the bound states of the electron in graphene quantum dots with a wedge disclination using density of states analysis are strongly dependent on the value of $n$. When changing the value of n, the positions as well as the amplitude and width size change. Positive values of $n$ and $\tau$ shift the peaks to the right. This is due to peak suppression. For $ n = 2 $ and $\tau=1$, we can see that only four peaks remain, with increasing amplitudes and decreasing width sizes (see Fig. \ref{fig01}e). In the case where $n$ takes a negative value and $\tau=1$, we observe that the resonance peaks change their initial positions and move to the left, which explains the appearance of three peaks corresponding to $ m = 5/2 $, $m = 11/2 $ and $m = 13/2 $ when $n=-2$ and $\tau=1$. Their amplitudes decrease as the size of the width increases. 
In the opposite case where $\tau = -1$, the effect of wedge disclination $n$ is reversed, such that for negative values of $n$ with $\tau = -1$, the resonance peaks move to the right with an increase in amplitude and a decrease in width. The condition \eqref{eq:resonanceposition} shows that there are three resonance peaks that are labeled according to their angular momentum $m=1/2$. However, they are missing for two cases: $(n=2, \tau=-1)$ and $(n=-2, \tau=1)$, as shown in  Fig. \ref{fig01}f,i. 
 
\begin{figure}[h]\centering 
 	\includegraphics[width=0.5\linewidth]{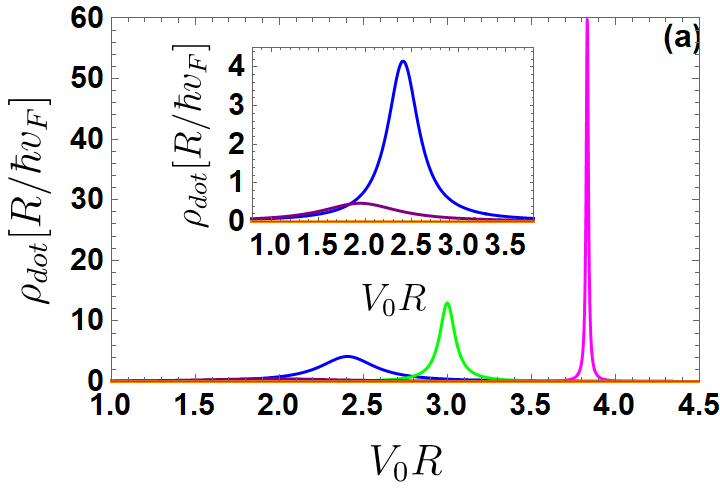}\includegraphics[width=0.5\linewidth]{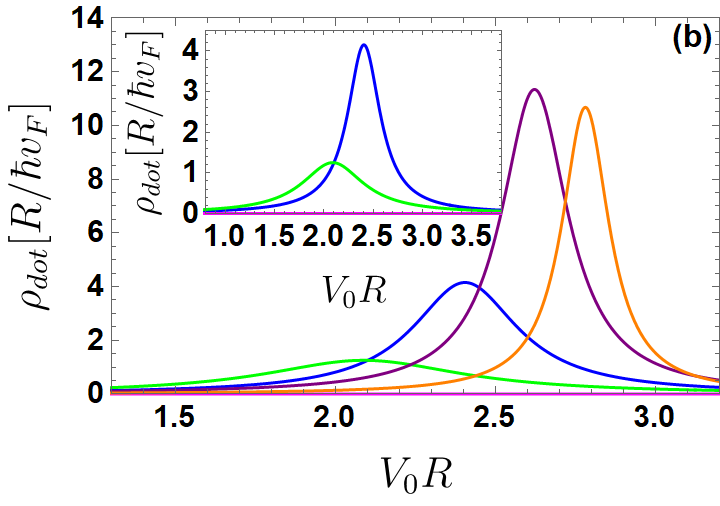}\\
 	\includegraphics[width=0.5\linewidth]{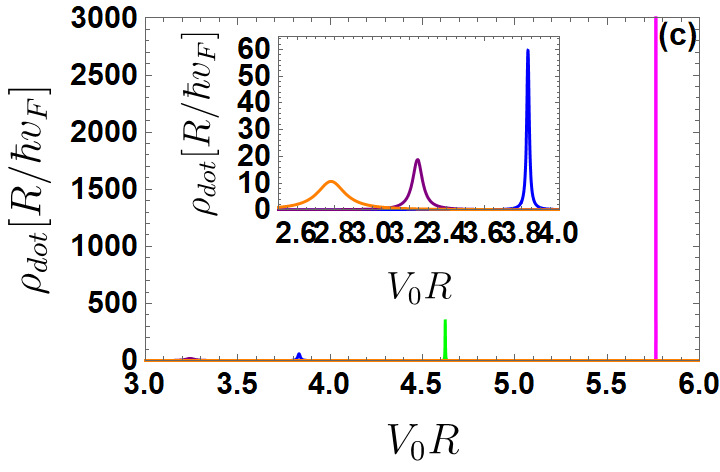}\includegraphics[width=0.5\linewidth]{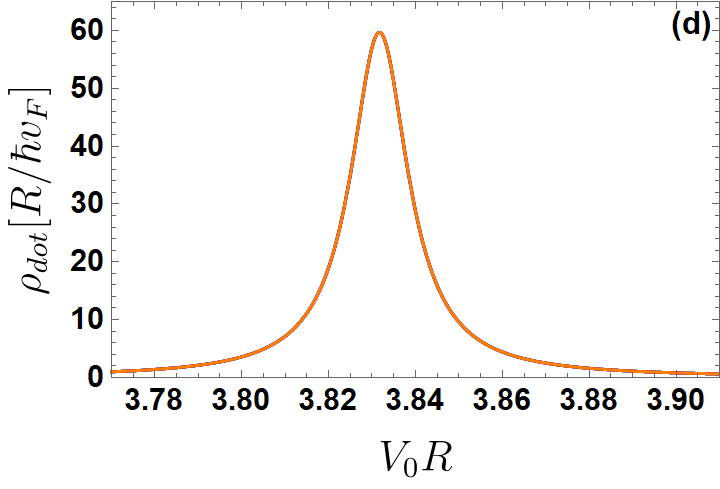}
 	\caption{(color online) 
 		DOS  as a function of gate voltage $ V_0R$ at $\epsilon =0$ with $R/L = 0.2$, $\Phi_i=0$ for  $n=0$ (blue line), $n=1$ (green line), $n=2$ (magenta line), $n=-1$ (purple line), $n=-2$ (orange line). (a, b):  $m=1/2$ and (c, d):  $m=3/2$ with (a, c): $\tau=1$  and (b, d): $\tau=-1$. }
 	\label{fig02}
 \end{figure}

 
%
The characteristics of the resonance are found to be dependent on the value of $n$ and the valley  $\tau$. The resonance positions shift to the right if $\tau n>0$ and become very straight and very important for $n = 2$ and $\tau=1$, whereas they shift to the left for $\tau n<0$. When $ n > 0 $, the influence of this magnitude where $n>0$  on the behavior of the resonance peaks is reversed. This result becomes very clear when  the DOS is plotted in Fig. \ref{fig02} for the first or the second resonance by varying the values of $n$ and $\tau$. 
In the case where $\tau=-1$, the parameter $n$ has no effect on the position or characteristics of the second resonance peak, as shown in Fig. \ref{fig02}d.
We note that the peaks of the second DOS resonance are very narrow compared to those of the first resonance. 
As  $n$ increases in Fig. \ref{fig02}c, the resonance peaks become very narrow, such that their amplitude increases but their width decreases.

\begin{figure}[h]\centering 
	\includegraphics[width=0.5\linewidth]{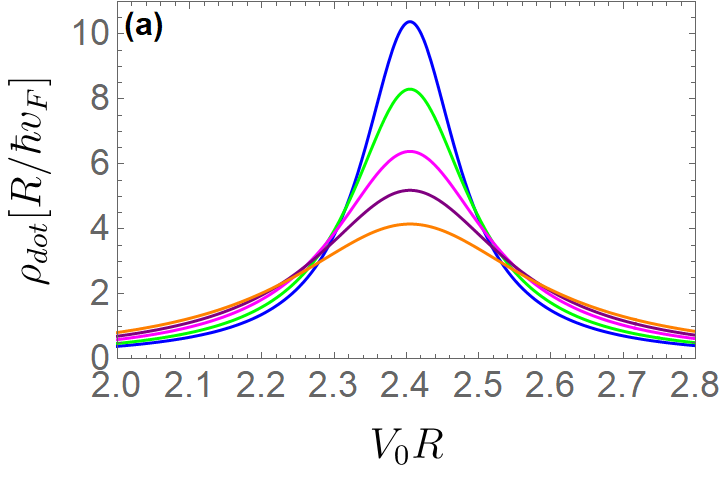}\includegraphics[width=0.5\linewidth]{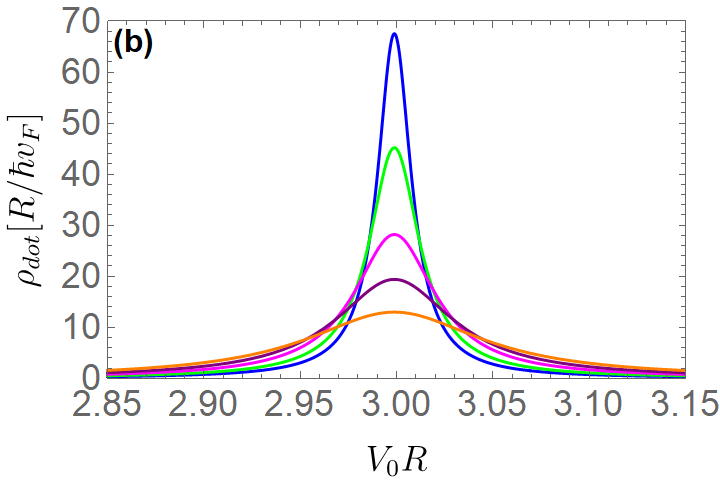}\\
	\includegraphics[width=0.5\linewidth]{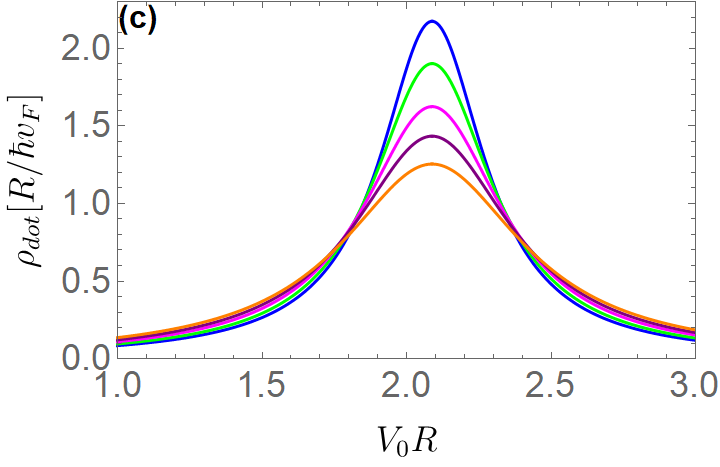}\includegraphics[width=0.515\linewidth]{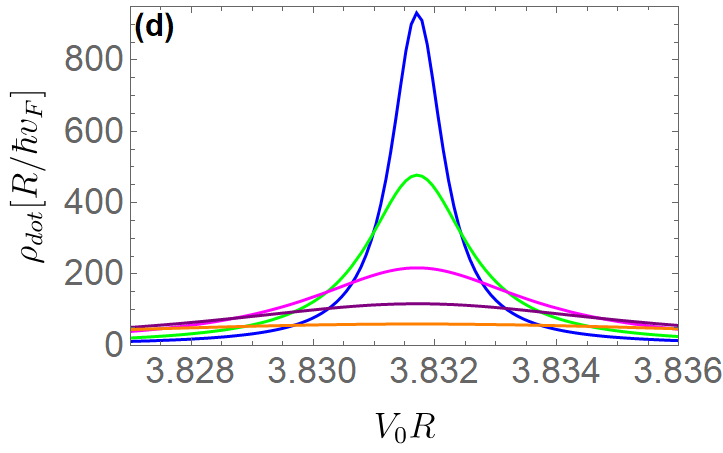}\\
	\includegraphics[width=0.5\linewidth]{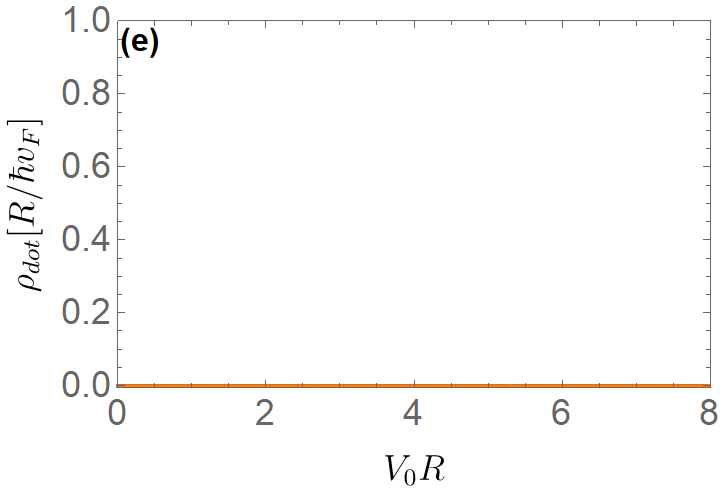}\includegraphics[width=0.505\linewidth]{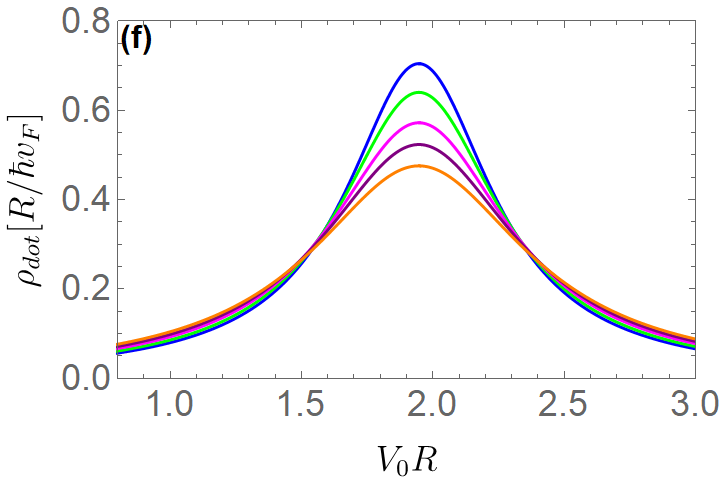}\\
\includegraphics[width=0.5\linewidth]{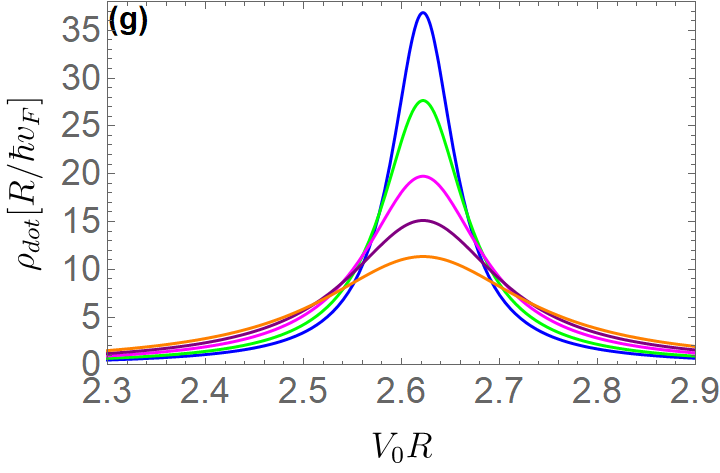}\includegraphics[width=0.515\linewidth]{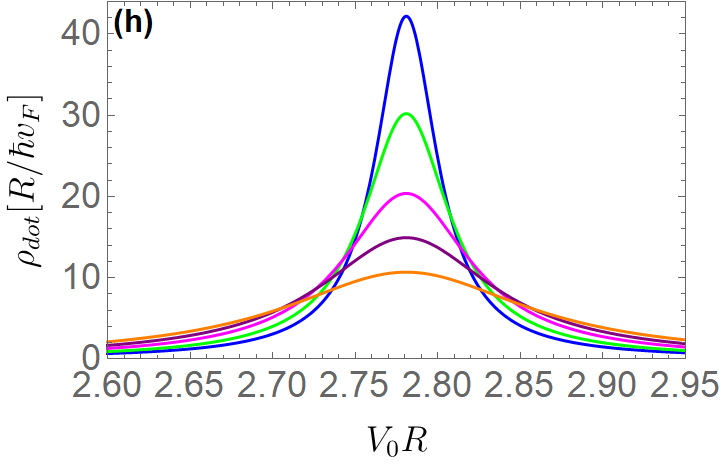}
	\caption{(color online) DOS  as a function of gate voltage $ V_0R$ at $\epsilon =0$  with $\Phi_i=0$ for  $m=1/2$ by changing the ratio $R/L=0.08$ (blue line), $R/L=0.1$ (green line), $R/L=0.13$ (magenta line), $R/L=0.16$ (purple line), $R/L=0.2$ (orange line). The pair $(n,\tau)$ is (a): $0$,  (b):  $(1,1)$, (c): $(1, -1)$,  (d):  $(2, 1)$, (e): $(\pm2,\mp 1)$,  (f): $(-1, 1)$, (g): $(-1, -1)$, (h): $(-2, -1)$.    }
	\label{fig03}
\end{figure}

In Fig. \ref{fig03}, we only consider the first resonance and investigate its influence on the magnitude of the point-lead coupling $R/L$, where $L$ denotes the distance between the source and the drain. 
We observe that the resonance peaks have Lorentzian shapes whose properties depend on several parameters, including the index $n$ and the ratio $R/L$. The resonance peaks become narrower when the quantum dot is very far from the source, i.e., when $R/L = 0$, as well as when the defect is chosen to be of square type, $n=2$, and for the valley $K$ ($\tau=1$). 
As a result, the DOS is a series of peaks influenced by the gate voltage $ V_0R$,  $n$, $R/L$, $m$, and $\tau$. 
The width of the resonances is independent of the coupling to the source and drain, which is determined by $R/L$, and the height increases when $R/L$ is very low, $n=2$, and $\tau=1$.  
 \begin{figure}[h]\centering 
 	\includegraphics[width=0.5\linewidth]{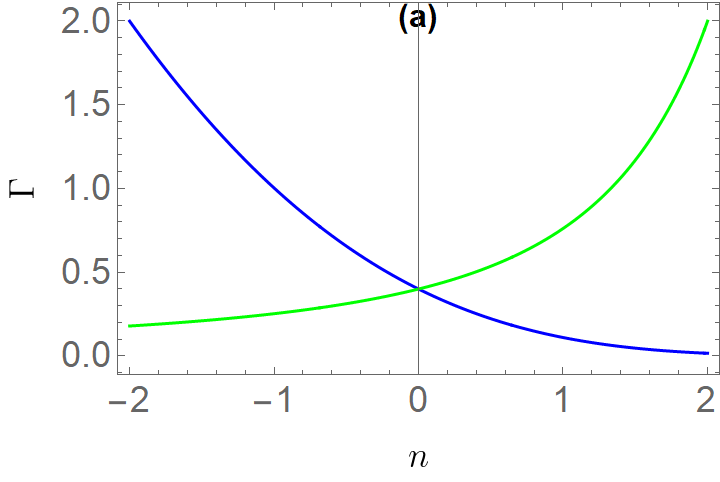}\includegraphics[width=0.5\linewidth]{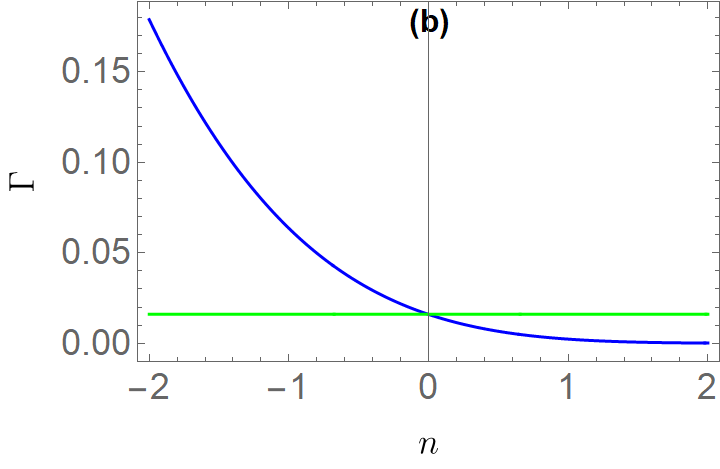}\\
 	\includegraphics[width=0.5\linewidth]{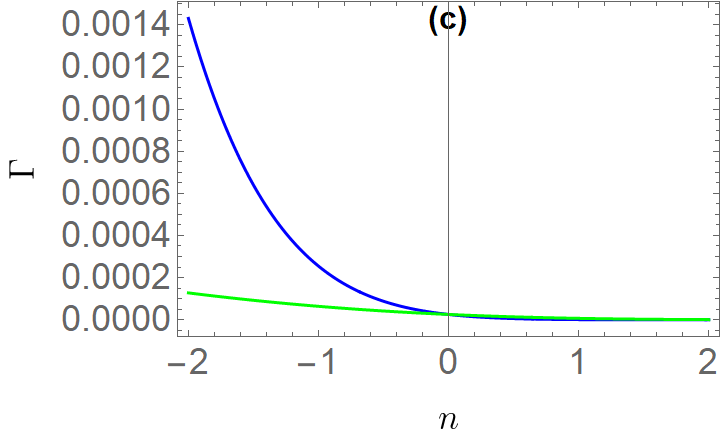}\includegraphics[width=0.5\linewidth]{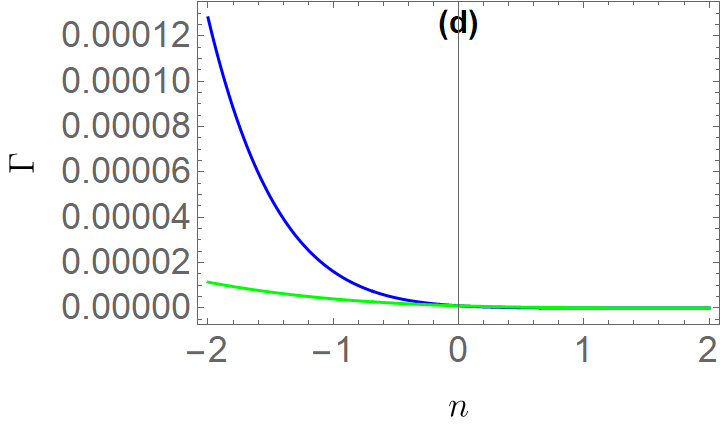}
 	\caption{(color online) The width of  resonance peak  as a function of wedge disclination $n$ at $\epsilon =0$ with $R/L = 0.2$, $\Phi_i=0$ for  (a): $m=1/2$, (b): $m=3/2$,   (c): $m=5/2$,   and (d):  $7=3/2$,   $\tau=1$ (blue line) and $\tau=-1$ (green line).}
 	\label{fig04}
 \end{figure}

 The width of the resonance peak as a function of wedge disclination $n$ is shown in Fig.~\ref{fig04} for  $m=1/2, 3/2, 5/2, 7/2$, and $\tau=\pm1$.
For higher values of   $m$ with $n=2$ and $ \tau=1$, one sees that the resonances become very sharp, which is consistent with the width scale \eqref{eq:resonanceposition}. 
The results presented in both Fig.~\ref{fig04}b and 
Fig.~\ref{fig02}d show that there is no influence of $n$ on the characteristics and position of the second resonance peak for $\tau=-1$. 
 On the other hand, the variation of the amplitude of resonance peaks is inversely related to the variation of their width. Consequently, these numerical values confirm the results obtained above.

\begin{figure}[h]\centering 
	\includegraphics[width=0.5\linewidth]{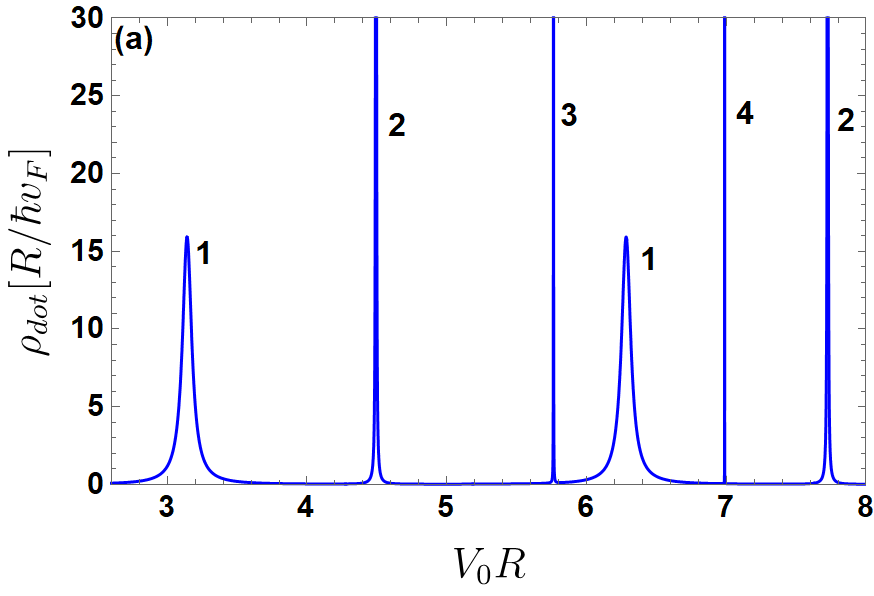}\includegraphics[width=0.5\linewidth]{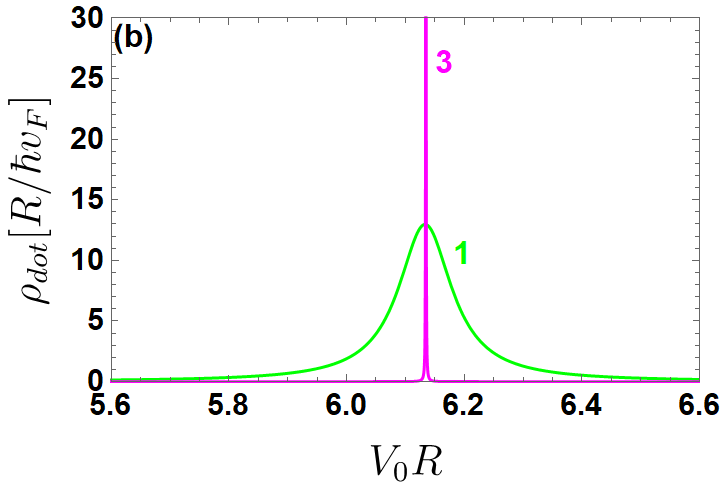}
	\includegraphics[width=0.5\linewidth]{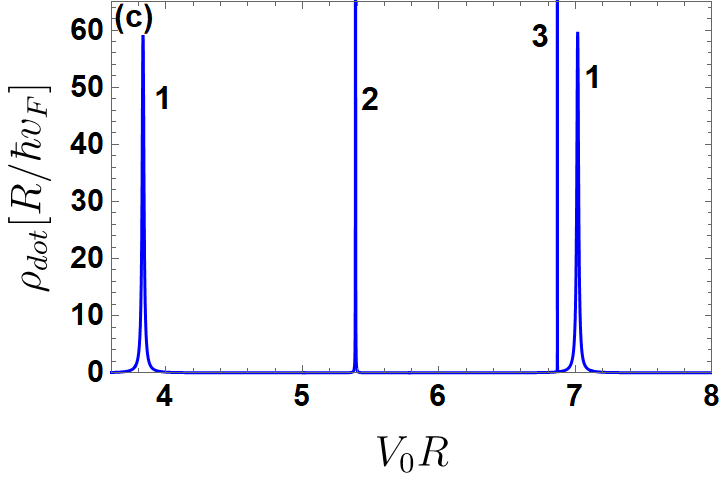}\includegraphics[width=0.5\linewidth]{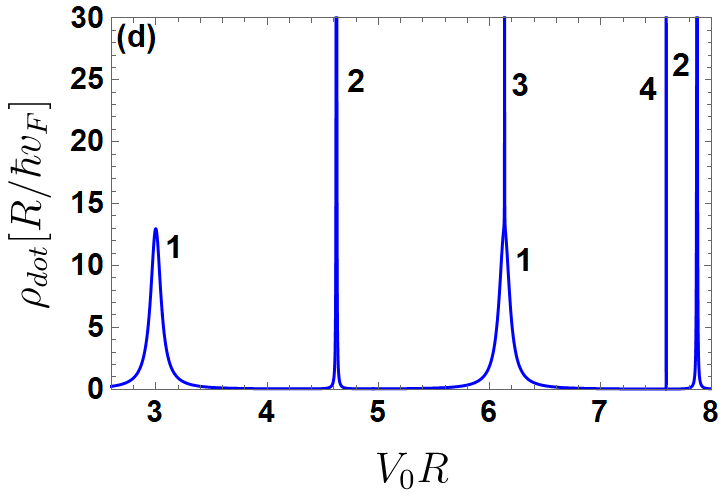} 
	\includegraphics[width=0.5\linewidth]{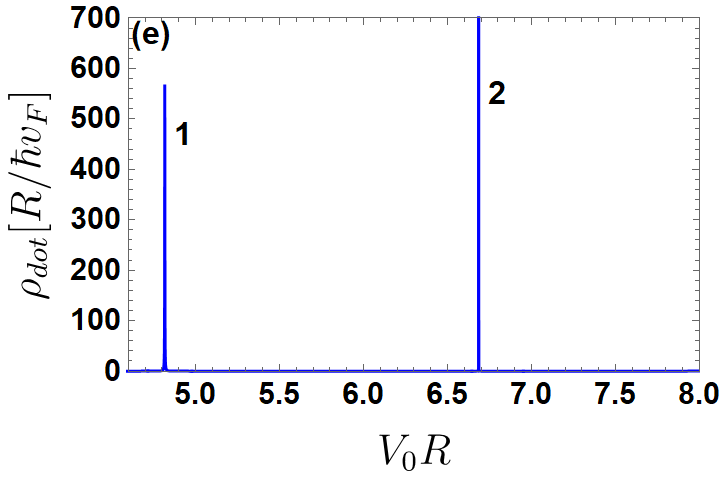}\includegraphics[width=0.5\linewidth]{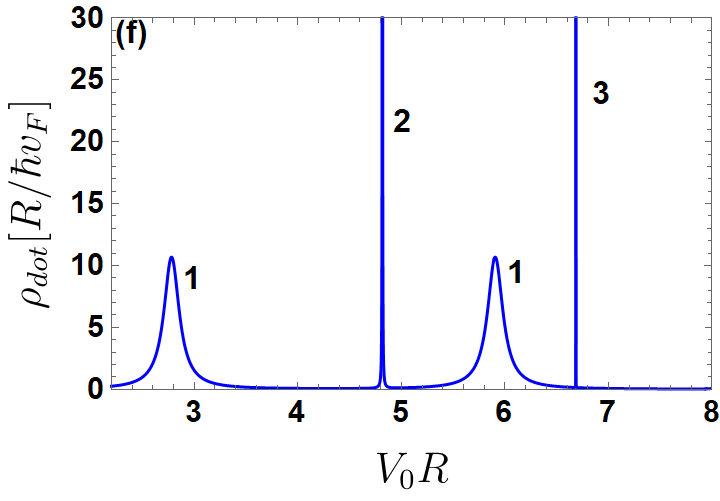}\\
	\includegraphics[width=0.5\linewidth]{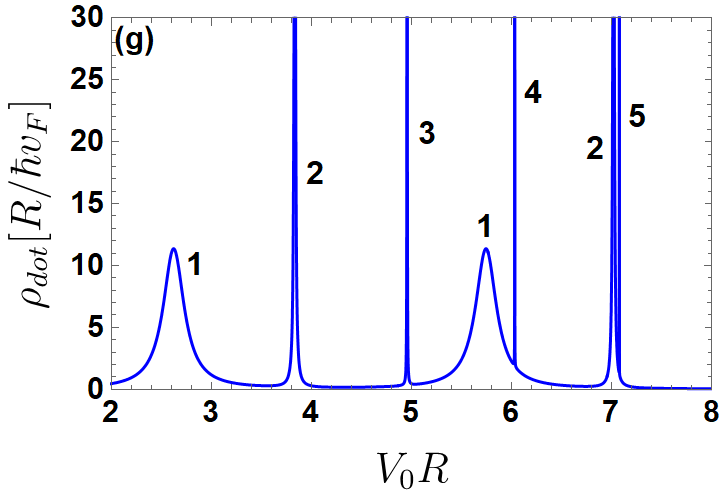}\includegraphics[width=0.5\linewidth]{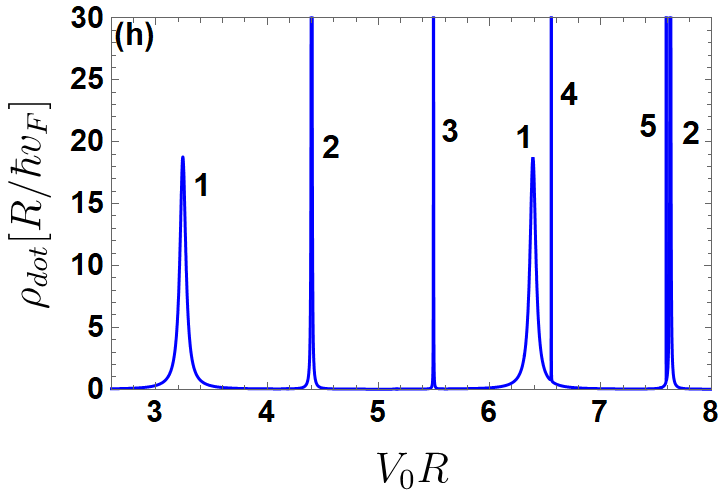}\\
	\includegraphics[width=0.5\linewidth]{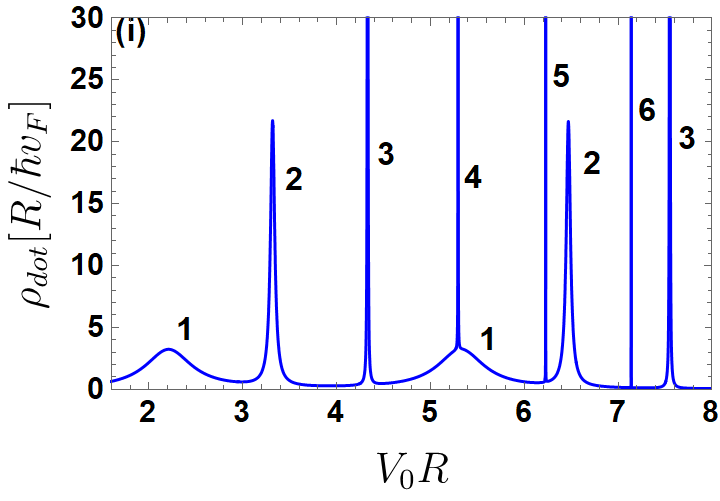}\includegraphics[width=0.5\linewidth]{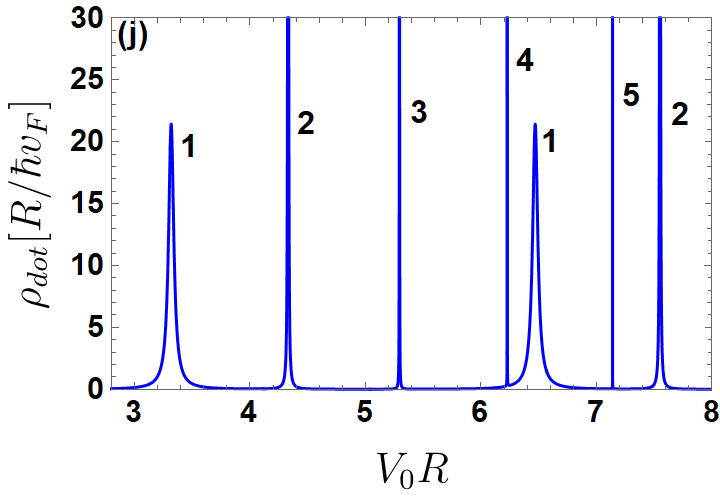}
	
	\caption{(color online) DOS  as a function of gate voltage $ V_0R$ at $\epsilon =0$ with $R/L = 0.2$, $\Phi_i=\frac{1}{2}$  for  $(n, \tau)$ such that (a): $0$, (c): $(1, 1)$, (d): $(1, -1)$, (e): $(2, 1)$, (f): $(2, -1)$, (g): $(-1, 1)$, (h): $(-1, -1)$, (i): $(-2, 1)$, (j): $(-2, -1)$. Here (b) is a zoom of (d) for $5.6<V_0 R<6.6$ and resonances are labeled by $|\mu|=1, \cdots , 6$.   }
	\label{fig05}
\end{figure} 
We now introduce a magnetic flux carried by a half magnetic flux of $\Phi_i=1/2$, yielding an integer total angular momentum denoted by $\mu$, which labels the resonance peaks. 
In Fig. \ref{fig05}, the DOS is shown as a function of the gate voltage $V_0 R$, with $R/L = 0.2$ for $n=0, \pm1, \pm2$, and $\tau=\pm1$. The electrons circling the flux tube collect an Aharonov-Bohm  phase $\pi$, canceling out the effect of the Berry phase, and the electron wave function collects via pseudo-rotation while moving in a circular motion. 
%
We find that when $\Phi_i=1/2$, the system enters a state with zero angular momentum, $\mu=0$, which cannot be confined by an electrostatic potential, even in the presence of a magnetic field.
We obtain results that are consistent with previous simulations \cite{bardarson2009electrostatic, titov2010charge, schneider2014density, bouhlal2021density} in terms of the position and scale of the width by changing the resonances $R/L$ when $ n = 0 $ in Fig. \ref{fig05}. Furthermore, one can observe extremely small resonance peaks in the presence of a magnetic flux.

\begin{figure}[h]\centering 
	\includegraphics[width=0.5\linewidth]{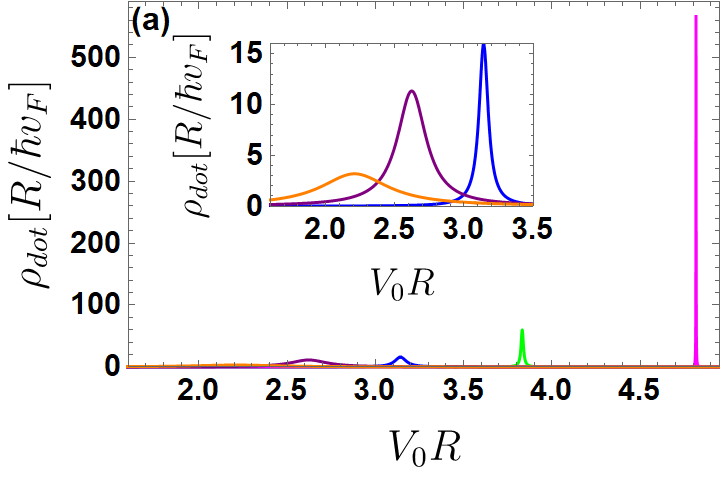}\includegraphics[width=0.5\linewidth]{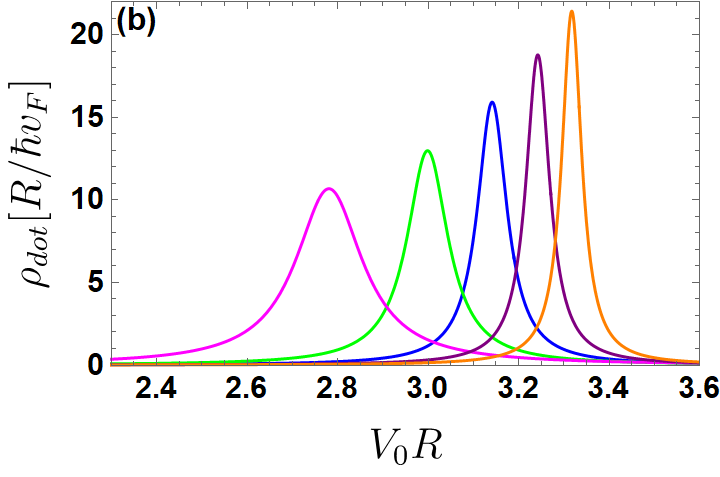}\\
	\includegraphics[width=0.5\linewidth]{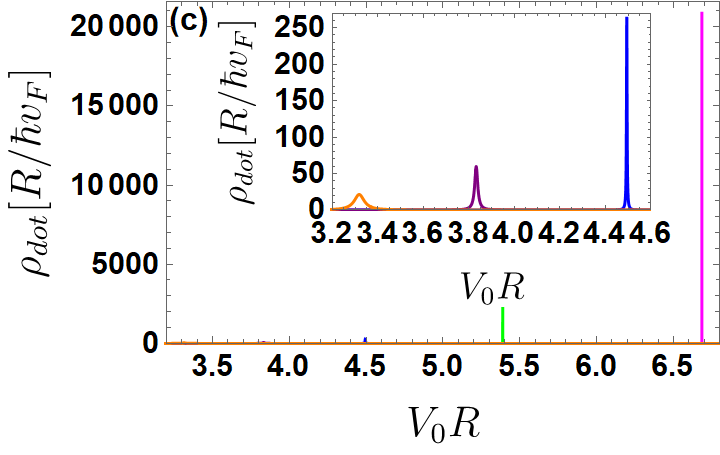}\includegraphics[width=0.5\linewidth]{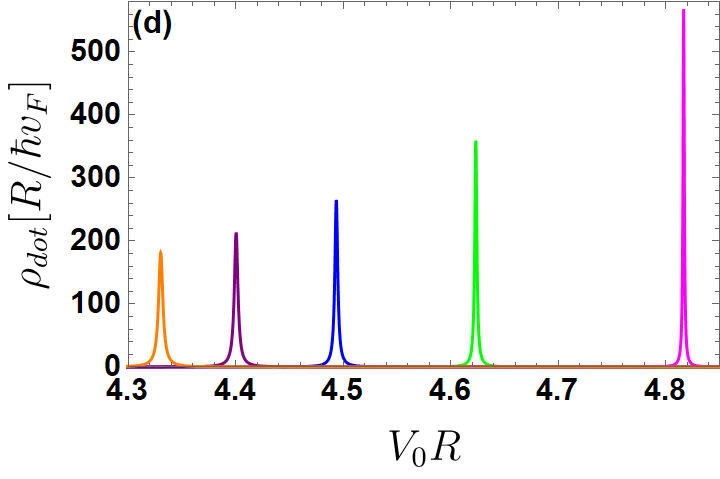}
	
	\caption{(color online) DOS  as a function of gate voltage $ V_0R$ at $\epsilon =0$ with $R/L = 0.2$, ${\Phi_i}=1/2$  for various wedge of disclination values: $n=0$ (blue line), $n=1$ (green line), $n=2$ (magenta line), $n=-1$ (purple line), $n=-2$ (orange line). (a, b):  $\mu=1$ and (c, d):  $\mu=2$ with (a, c): $\tau=1$  and (b, d): $\tau=-1$.}
	\label{fig06}
\end{figure}

For $\tau =\pm1$, the DOS of the first and second resonances $\mu = 1,2$ is represented in Fig. \ref{fig06} by varying the value of the wedge disclination $n = 0$ (blue line), $n = 1$ (green line), $n = 2$ (magenta line), $n=-1$ (purple line), and $n=-2$ (orange line). 

 \begin{figure}[h]\centering 
 	\includegraphics[width=0.5\linewidth]{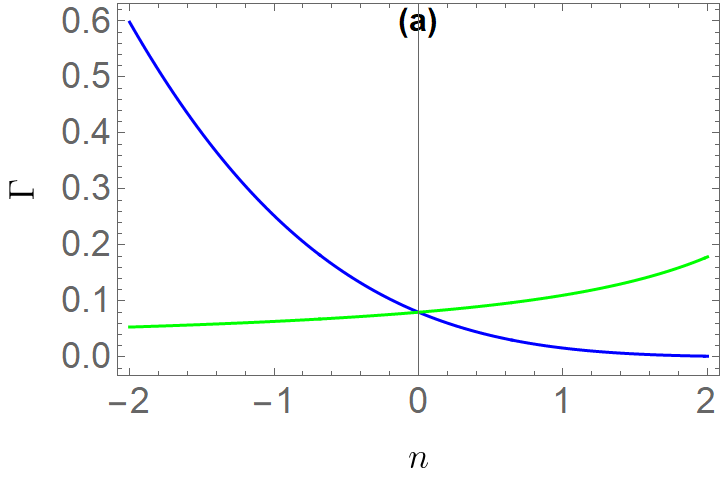}\includegraphics[width=0.5\linewidth]{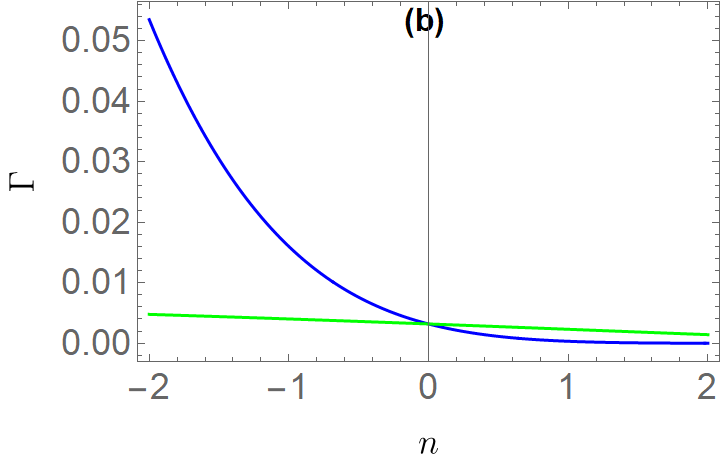}\\
 	\includegraphics[width=0.5\linewidth]{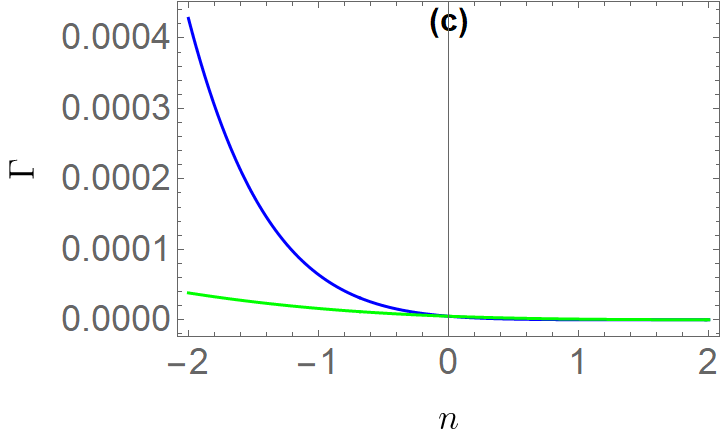}\includegraphics[width=0.5\linewidth]{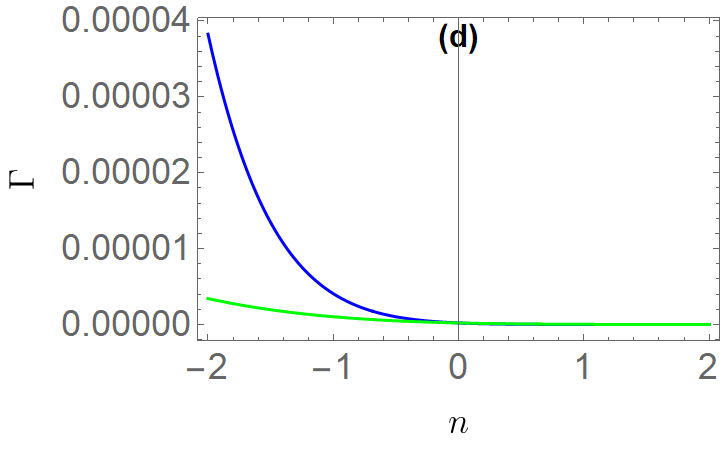}
 	\caption{(color online) The width of  the resonance peak  as a function of wedge disclination $n$ at $\epsilon =0$ with $R/L = 0.2$, $\Phi_i=\frac{1}{2}$ for   (a): $\mu=1$, (b): $\mu=2$, (c): $\mu=3$ and (d): $\mu=4$  such that $\tau=1$ (blue line) and $\tau=-1$ (green line).}
 	\label{fig08}
 \end{figure}
We notice that the effect of a wedge disclination becomes very important, as shown in Fig. \ref{fig05}e for $n=2$ and $\tau=1$. We  find very narrow resonance peaks well shifted to the right, which leads to a suppression of the four peaks. 
For $(n =-2, \tau=1$ and under the influence of a magnetic flux, however, we notice a decrease in the amplitude of the peaks as well as the appearance of three peaks, indicating that the number of bound states increases, see Fig. \ref{fig05}. 
%
%

\begin{figure}[h]\centering 
	\includegraphics[width=0.5\linewidth]{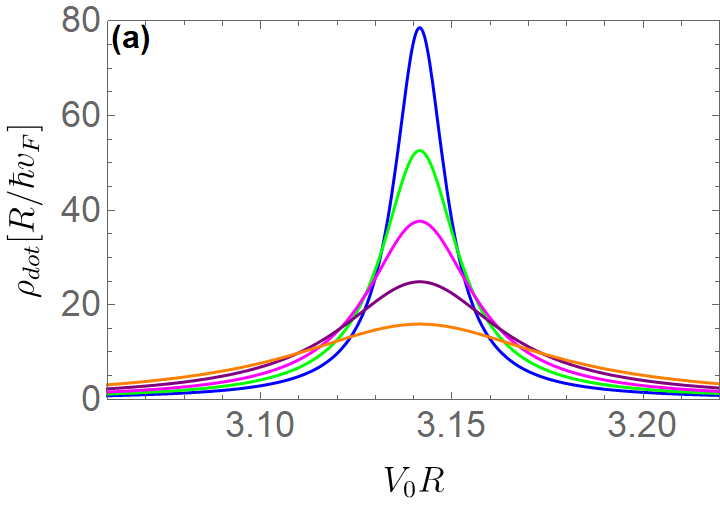}\\
	\includegraphics[width=0.5\linewidth]{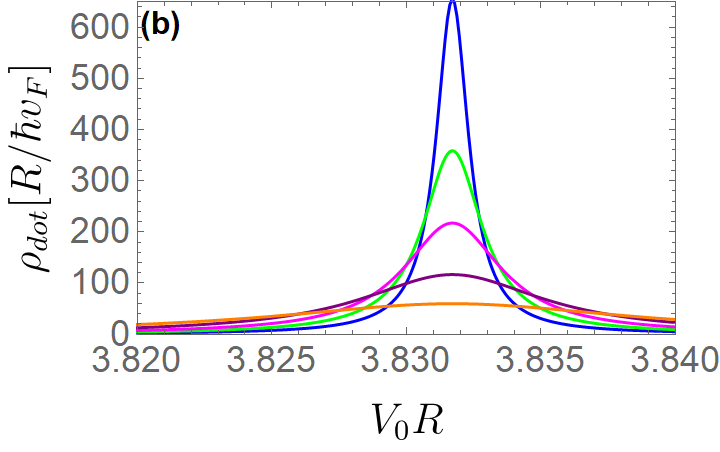}\includegraphics[width=0.485\linewidth]{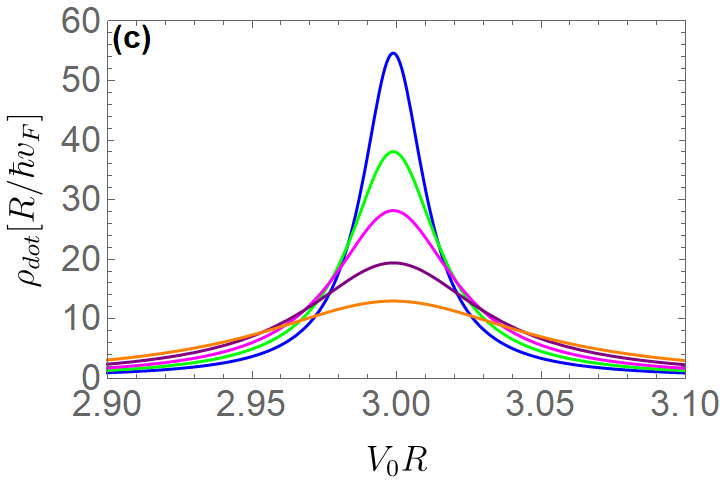}\\
	\includegraphics[width=0.495\linewidth]{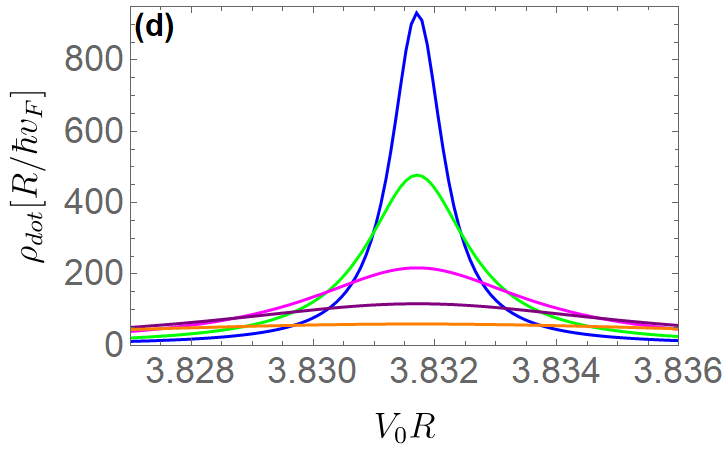}\includegraphics[width=0.475\linewidth]{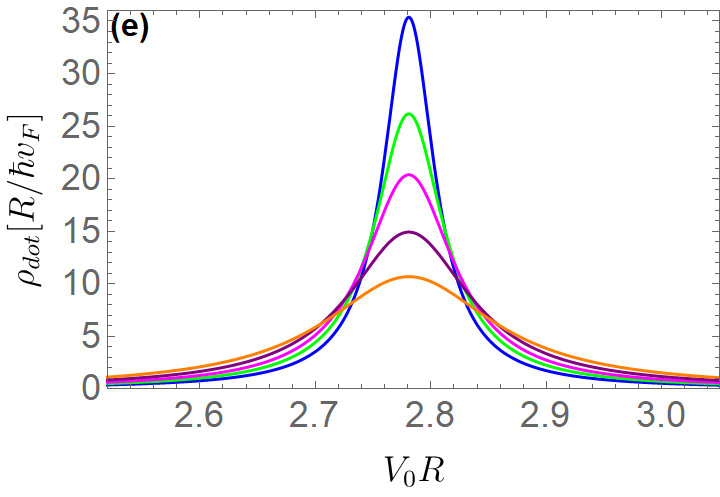}\\
   \includegraphics[width=0.49\linewidth]{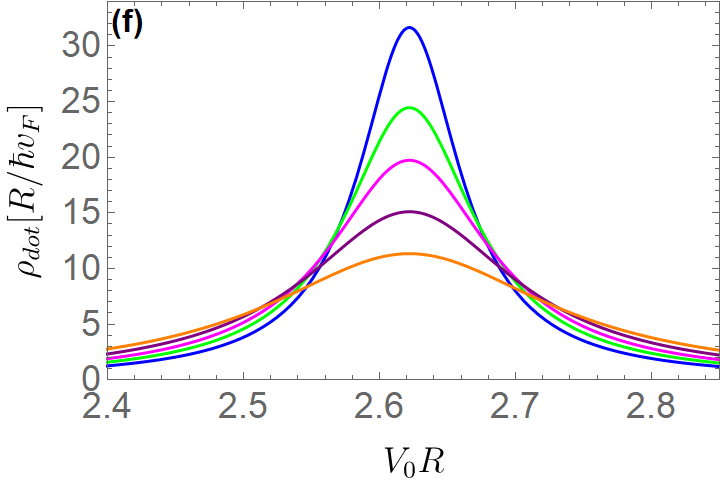}\includegraphics[width=0.515\linewidth]{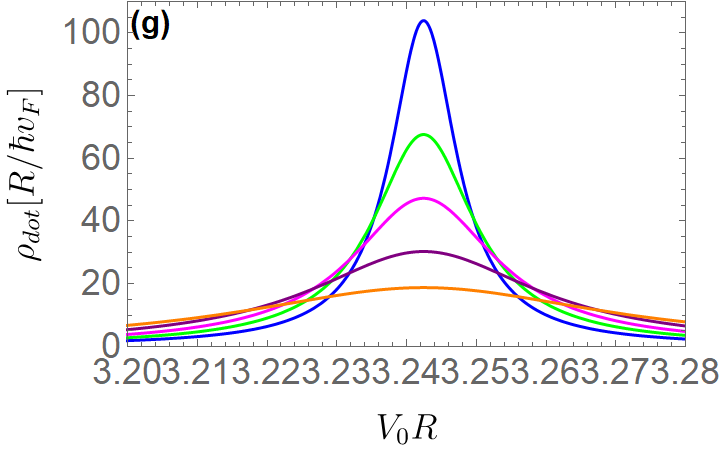}\\
	\includegraphics[width=0.5\linewidth]{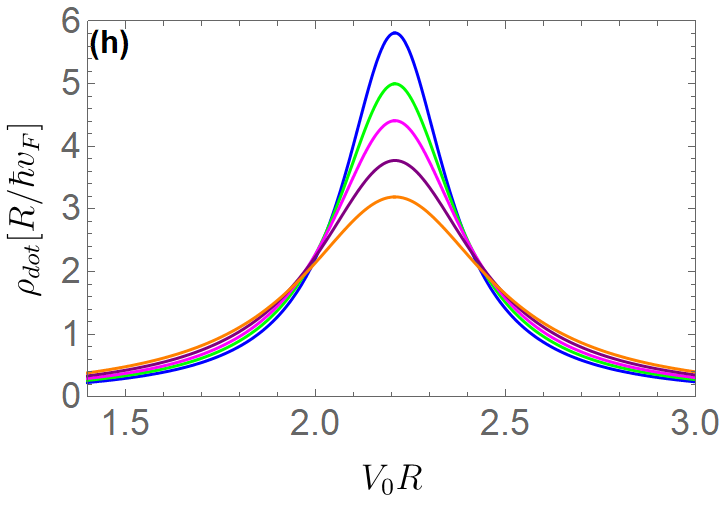}\includegraphics[width=0.515\linewidth]{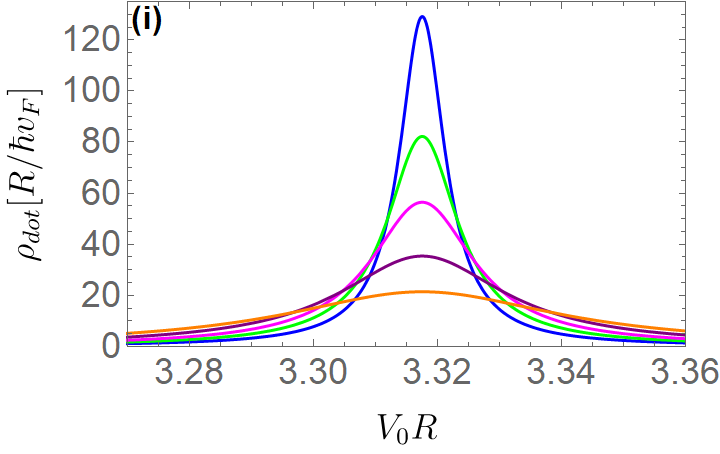}
	\caption{(color online) DOS  as a function of gate voltage $ V_0R$ at $\epsilon =0$ with $\Phi_i=1/2$ and for the first resonance $\mu=1$ by changing the  ratio $R/L=0.08$ (blue line), $R/L=0.1$ (green line), $R/L=0.13$ (magenta line), $R/L=0.16$ (purple line), $R/L=0.2$ (orange line). With the pair $ (n,\tau) $ such that (a): 0,  (b):  $(1, 1)$, (c): $(1, -1)$,  (d):  $(2, 1)$, (e): $(2, -1)$, (f): $(-1, 1)$, (g): $(-1, -1)$, (h): $(-2, 1)$, (i): $(-2, -1)$.}
	\label{fig07}
\end{figure}

 \begin{figure}[h]\centering 
	\includegraphics[width=0.5\linewidth]{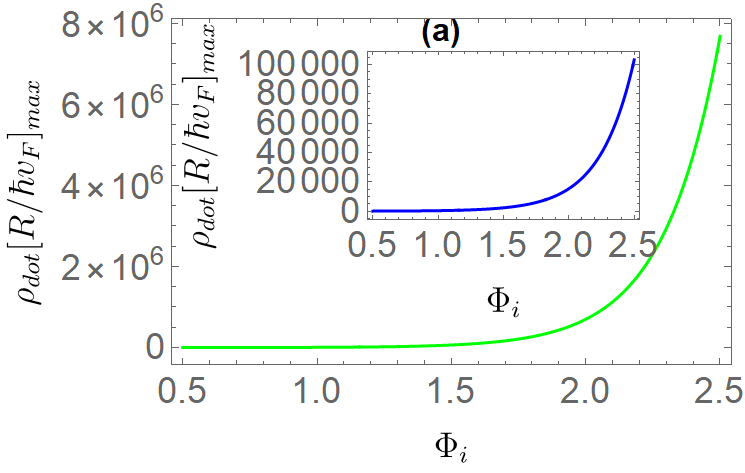}\includegraphics[width=0.5\linewidth]{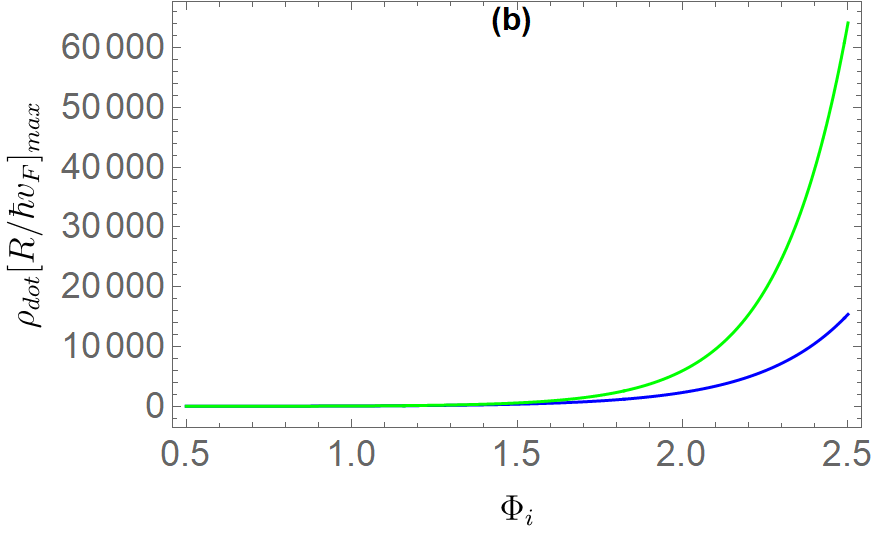}
	
	\caption{(color online) Amplitude of  the resonance peak as function of the magnetic flux  $\Phi_i$ at  $\epsilon =0$ with  $R/L = 0.2$ for  $n=1$ (blue line), $n=2$ (green line) and $V_0^{\prime} R=V_0 R$.  }
	\label{fig09}
\end{figure}
We investigate the effect of the ratio $R/L$ on DOS for the first resonance $\mu =1$ in the presence of a magnetic flux $\Phi_i=1/2$ with $\tau=\pm1$ by varying the value of a wedge disclination $n$. 
By comparing the results of the two Figs. \ref{fig03} and \ref{fig07}, we show that the analysis can be made quantitative by taking a specific peak and studying the effect of the ratio $R/L$.  We observe that DOS becomes very saturated when the limit $R/L \to 0$, which corresponds to a weak coupling between the quantum dot and a ring-shaped metallic. 
Finally, the behavior of the resonance peaks is affected by three physical quantities, which are the wedge disclination value $n$, the magnetic flux $\Phi_i=1/2$, and the ratio $R/L$. To give a better presentation of the result, we draw another curve that represents the width of the resonance peaks in the presence of magnetic flux $\Phi_i=1/2$.
%
%


The width of the resonance peak for $\mu= 1, 2, 3, 4$ is plotted in Fig. \ref{fig08} for two values of $\tau = +1 (-1)$. Comparing the results of the two Fig. \ref{fig04} and Fig. \ref{fig08}, we see that the wedge disclination effect becomes very important for the valley $K (\tau=1)$. Also, we notice that its effect enjoys the presence of a magnetic flux and for $\tau=1$, because it is better.

The amplitude of a DOS resonance peak as a function of magnetic flux $\Phi_i$ for the angular momentum $m=1/2 $ and wedge disclination $n=1, 2$ is shown in Fig. \ref{fig09}. 
We see that the amplitude of a resonance peak grows exponentially as a function of the magnetic flux and the variation becomes very important when the value of $n$ is fixed at the value $2$ and for $\tau=1$. This result is very similar to the previous results obtained in \cite{bouhlal2021density, gutierrez2015mass}. 
We conclude that the wedge disclination $n$ is a physical quantity that can alter the behavior of the resonance peaks that make up the DOS. Furthermore, its effect is amplified in the presence of a magnetic flux. As a result, we can say that it can be used as a tunable physical parameter for the control of the transport properties of our system.

\section{Conclusion} \label{cc}


We considered a model of graphene quantum dot in the presence of magnetic flux $\Phi_i$ undergoing a default, which can be understood from Volterra's constructions. To treat the problem properly, we started by solving the corresponding Dirac equation and analytically determining the energy spectra, taking into account the cylindrical symmetry. We found that the solutions are expressed in terms of Hankel functions. Then, we obtained an approximate formula for the density of states (DOS) as a function of physical parameters such as magnetic flux, angular momentum, wedge disclination, and applied electrostatic potential.


The analysis of the results shows that even if graphene does not present a gap in its electronic energy spectrum, we can have the confinement of electrons under the effect of an external electrostatic excitation in a quantum dot surrounded by a large sheet of undoped graphene. We have shown that the density of states is an alternative technique that can be used to study the phenomenon of fermions confinement in a graphene quantum dot. We have performed an analysis that considers the integrated density of states on the dot and an undoped graphene sheet. The formalism we have followed in this paper can easily be extended to this type of measurement configuration. However, with respect to the qualitative analysis of the width, amplitude, and position of the peaks as well as the suppression and creation of new resonance peaks. 

We presented our numerical results in terms of the $R/L$ ratio, $\Phi_i$ magnetic flux, and  wedge disclination $n$.
 We have shown that the presence of wedge disclination modifies the behavior of the resonance peaks and also changes their positions. We found that for negative values of $n$, they can create new resonance peaks and minimize the resonance phenomenon by decreasing the amplitude of the peaks and increasing their widths, thus increasing the number of bound electron states. On the other hand, we have shown that the result of adding a positive $n$ term that represents the defect is a narrow peak that becomes very narrow in the presence of magnetic flux as well as the suppression of some peaks. Finally, we discovered that the density of states is very divergent in the presence of a very strong magnetic field and large values of $ n = 2 $ (square defect). These results could have potential applications in nanoelectronics.



\bibliography{ref.bib}

\end{document}